\overfullrule=0pt
\input harvmac
\def\a{{\alpha}}

\def\l{{\lambda}}
\def\lb{{\overline\lambda}}
\def\b{{\beta}}

\def\g{{\gamma}}

\def\om{{\omega}}
\def\d{{\delta}}
\def\e{{\epsilon}}
\def\s{{\sigma}}
\def\k{{\kappa}}

\def\half{{1\over 2}}
\def\p{{\partial}}
\def\pb{{\overline\partial}}
\def\t{{\theta}}
\def\tb{{\overline\theta}}
\def\bar{\overline}

\Title{\vbox{\hbox{IFT-P.020/2004 }}}
{\vbox{
\centerline{\bf Multiloop Amplitudes and Vanishing Theorems}
\centerline{\bf using the Pure Spinor
Formalism for the Superstring}}}
\bigskip\centerline{Nathan Berkovits\foot{e-mail: nberkovi@ift.unesp.br}}
\bigskip
\centerline{\it Instituto de F\'\i sica Te\'orica, Universidade Estadual
Paulista}
\centerline{\it Rua Pamplona 145, 01405-900, S\~ao Paulo, SP, Brasil}

\vskip .3in
A ten-dimensional super-Poincar\'e covariant formalism
for the superstring was recently developed which involves a BRST operator 
constructed from superspace matter variables and a pure spinor ghost
variable. A super-Poincar\'e covariant prescription was defined for
computing tree amplitudes and was shown to coincide with the 
standard RNS prescription. 

In this paper, picture-changing operators are used to
define functional integration over the pure spinor ghosts and 
and to construct a suitable $b$ ghost. A super-Poincar\'e covariant
prescription is then given for the computation of $N$-point multiloop
amplitudes. 
One can easily prove that massless $N$-point 
multiloop amplitudes vanish for $N<4$, confirming the perturbative finiteness
of superstring theory. One can also prove the Type IIB S-duality
conjecture that $R^4$ terms in the effective action receive no
perturbative contributions above one loop.

\vskip .3in

\Date {June 2004}

\newsec{Introduction}

The computation of multiloop amplitudes in superstring theory
has many important applications such as verifying perturbative finiteness
and testing duality conjectures. Nevertheless, this subject has 
received little attention over the last fifteen years, mainly because
of difficulties in computing multiloop amplitudes using either the 
Ramond-Neveu-Schwarz (RNS) or Green-Schwarz (GS) formalism. 

In the
RNS formalism, spacetime supersymmetric amplitudes are obtained 
after summing over spin structures, which can be done explicitly only
when the number of loops and external states is small \ref\parkes
{O. Lechtenfeld and A. Parkes,
{\it On Covariant Multiloop Superstring Amplitudes},
Nucl. Phys. B332 (1990) 39.}.
Since there are divergences near the boundary of moduli space before
summing over spin structures, surface terms in the amplitude expressions
need to be treated with care \ref\verlone{E. Verlinde and H. Verlinde,
{\it Multiloop Calculations in Covariant Superstring Theory}, Phys. Lett.
B192 (1987) 95.}\ref\verltwo{
H. Verlinde, {\it The Path Integral Formulation of
Supersymmetric String Theory}, PhD Thesis, Univ. of Utrecht (1988)\semi
H. Verlinde,
{\it A Note on the Integral over Fermionic Supermoduli},
Utrecht Preprint No. THU-87/26 (1987) unpublished.}
\ref\atick{J. Atick, G. Moore, and A. Sen,
{\it Some Global Issues in String Perturbation Theory},
Nucl. Phys. B308 (1988) 1\semi
J. Atick, G. Moore and A. Sen,
{\it Catoptric Tadpoles},
Nucl. Phys. B307 (1988) 221\semi
J. Atick, J. Rabin and A. Sen,
{\it An Ambiguity in Fermionic String Perturbation Theory},
Nucl. Phys. B299 (1988) 279.}
\ref\phong{E. D'Hoker and D.H. Phong, {\it Two Loop Superstrings, 
1. Main Formulas}, Phys. Lett. B529 (2002)
241, hep-th/0110247\semi
E. D'Hoker and D.H. Phong,
{\it Two Loop Superstrings, 2. The Chiral Measure on Moduli Space}, Nucl.
Phys. B636 (2002) 3, hep-th/0110283\semi
E. D'Hoker and D.H. Phong,
{\it Two Loop Superstrings, 3. Slice Independence and Absence
of Ambiguities}, Nucl. Phys. B636 (2002) 61, hep-th/0111016\semi
E. D'Hoker and D.H. Phong,
{\it Two Loop Superstrings, 4. The Cosmological Constant and Modular
Forms},
Nucl. Phys. B639 (2002) 129, hep-th/0111040.}. 
Furthermore, the complicated nature
of the Ramond vertex operator in the RNS formalism 
\ref\fms{D. Friedan, E. Martinec and S. Shenker, {\it Conformal Invariance,
Supersymmetry and String Theory}, Nucl. Phys. B271 (1986) 93.}
makes it difficult
to compute amplitudes involving external fermions or Ramond-Ramond bosons.
For these reasons, up to now, explicit multiloop computations in the RNS
formalism have been limited to four-point two-loop amplitudes involving
external Neveu-Schwarz bosons \ref\iengo{R. Iengo and C.-J. Zhu, {\it
Two Loop Computation of the Four-Particle
Amplitude in the Heterotic String}, Phys. Lett. B212 (1988) 313\semi
R. Iengo and C.-J. Zhu, {\it Explicit Modular Invariant Two-Loop Superstring
Amplitude Relevant for $R^4$}, JHEP 06 (1999) 011, hep-th/9905050
\semi R. Iengo, {\it
Computing the $R^4$ Term at Two Superstring Loops}, JHEP 0202 (2002) 035,
hep-th/0202058\semi
Z.-J. Zheng, J.-B. Wu and C.-J. Zhu, {\it
Two-Loop Superstrings in Hyperelliptic Language I: the Main
Results}, Phys. Lett. B559 (2003) 89, hep-th/0212191\semi
Z.-J. Zheng, J.-B. Wu and C.-J. Zhu, {\it
Two-Loop Superstrings in Hyperelliptic Language II: the Vanishing
of the Cosmological Constant and the Non-Renormalization Theorem},
Nucl. Phys. B663 (2003) 79, hep-th/0212198\semi
Z.-J. Zheng, J.-B. Wu and C.-J. Zhu, {\it
Two-Loop Superstrings in Hyperelliptic Language III: the Four-Particle
Amplitude}, Nucl. Phys. B663 (2003) 95, hep-th/0212219\semi
W.-J. Bao and C.-J. Zhu, {\it Comments on Two-Loop Four-Particle
Amplitude in Superstring Theory}, JHEP 0305 (2003) 056, hep-th/0303152.
}\phong.\foot{Danilov \ref\danilov{G.S. Danilov,
{\it Finiteness of Multiloop Superstring Amplitudes}, hep-th/9801013\semi
G.S. Danilov, {\it On Cruel Mistakes in the Calculation of Multiloop Superstring
Amplitudes}, hep-th/0312177.} has claimed to be able
to compute RNS amplitudes for arbitrary genus, however, this author
has been unable to understand his methods.} 

In the GS formalism, spacetime supersymmetry is manifest but one needs
to fix light-cone gauge and introduce non-covariant operators at the
interaction points of the Mandelstam string diagram\ref\GS{M.B. Green
and J.H. Schwarz, {\it Superstring Interactions}, Nucl. Phys. B218 (1983)
43\semi M.B. Green and J.H. Schwarz,
{\it Superstring Field Theory}, Nucl. Phys. B243
(1984) 475.}\ref\mandelstam{S. Mandelstam,
{\it Interacting String Picture of the
Neveu-Schwarz-Ramond Model}, Nucl. Phys. B69 (1974) 77.}\ref\mandelstamtwo{
S. Mandelstam, {\it Interacting String Picture of the Fermionic String},
Prog. Theor. Phys. Suppl. 86 (1986) 163.}. Because of
complications caused by these non-covariant interaction point operators
\ref\greensite{J. Greensite and F.R. Klinkhamer, {\it Superstring 
Amplitudes and Contact Interactions}, Nucl. Phys. B304 (1988) 108\semi
M.B. Green and N. Seiberg, {\it Contact Interactions in Superstring
Theory}, Nucl. Phys. B299 (1988) 559.},
explicit amplitude expressions have been computed using the light-cone GS
formalism only for four-point tree and one-loop amplitudes \GS.\foot{
Although multiloop GS expressions were obtained by Restuccia and
Taylor in
\ref\rt{A. Restuccia and J.G. Taylor,
{\it The Construction of Multiloop Superstring Amplitudes in the Light
Cone Gauge}, Phys. Rev. D36 (1987) 489\semi
A. Retuccia and J.G. Taylor, {\it
Light-cone Gauge Analysis of Superstrings}, Physics Reports 174 (1989) 283.},
this author does not think that they correctly took 
into account the
contact terms between interaction-point operators. 
Note that
the $N$-point tree amplitudes proposed
by Mandelstam in \mandelstamtwo\ were derived using unitarity arguments
and were not directly computed from the GS formalism.}

Four years ago, a new formalism for the superstring was proposed
\ref\superp{N. Berkovits, {\it Super-Poincar\'e Covariant Quantization of the
Superstring}, JHEP 04 (2000) 018, hep-th/0001035.}\ref\ictp{N. Berkovits,
{\it ICTP Lectures on Covariant Quantization of the
Superstring}, hep-th/0209059.}
with manifest ten-dimensional super-Poincar\'e covariance.
In conformal gauge, the worldsheet action is quadratic and physical
states are defined using a BRST operator constructed from superspace
matter variables and a pure spinor ghost variable. A super-Poincar\'e
covariant prescription was given for computing $N$-point tree amplitudes,
which was later shown to coincide with the standard RNS prescription \ref
\vallilo{N. Berkovits and B.C. Vallilo, {\it
Consistency of Super-Poincar\'e Covariant Superstring
Tree Amplitudes}, JHEP 07 (2000) 015, hep-th/0004171.}\ref\relating
{N. Berkovits, {\it Relating the RNS and Pure Spinor Formalisms for the
Superstring}, JHEP 08 (2001) 026, hep-th/0104247.}.
It was also proven that the BRST cohomology reproduces the correct
superstring spectrum \ref\cohom{N. Berkovits, {\it
Cohomology in the Pure Spinor Formalism for the
Superstring}, JHEP 09 (2000) 046, hep-th/0006003\semi
N. Berkovits and O. Chand\'{\i}a,
{\it Lorentz Invariance of the Pure Spinor
BRST Cohomology for the Superstring}, 
Phys. Lett. B514 (2001) 394, hep-th/0105149.}
and that BRST invariance in a curved supergravity
background implies the low-energy superspace
equations of motion for the background superfields \ref\howeme{N. Berkovits
and P. Howe, {\it Ten-Dimensional Supergravity Constraints from the
Pure Spinor Formalism for the Superstring}
Nucl. Phys. B635 (2002) 75, hep-th/0112160.}\ref\vchan{B.C. Vallilo and O.
Chand\'{\i}a, {\it Conformal Invariance of the Pure Spinor Superstring
in a Curved Background}, JHEP 0404 (2004) 041, hep-th/0401226.}. 

Because of the pure spinor constraint satisfied by the worldsheet ghosts,
it was not known how to define functional integration in this formalism.
For this reason, the tree amplitude prescription in 
\superp\ relied on BRST cohomology for defining the correct normalization
of the worldsheet zero modes. Furthermore, there was no natural $b$ ghost
in this formalism, which made it difficult to define amplitudes
in a worldsheet
reparameterization-invariant manner. Because of these complications,
it was not clear how to compute loop amplitudes using this formalism and
other groups looked for ways of relaxing the pure spinor constraint
without modifying the BRST cohomology \ref\vann{P.A. Grassi, G. Policastro,
M. Porrati and P. Van Nieuwenhuizen, {\it Covariant Quantization of 
Superstrings without Pure Spinor Constraints}, JHEP 0210 (2002) 054,
hep-th/0112162\semi 
P.A. Grassi, G. Policastro and
P. Van Nieuwenhuizen, {\it The Massless Spectrum of Covariant  
Superstrings}, JHEP 0211 (2002) 001,
hep-th/0202123\semi 
P.A. Grassi, G. Policastro and
P. Van Nieuwenhuizen, 
{\it On the BRST Cohomology of Superstrings with/without Pure Spinors}
Adv. Theor. Math. Phys. 7 (2003) 499, hep-th/0206216\semi
P.A. Grassi, G. Policastro and
P. Van Nieuwenhuizen, 
{\it An Introduction to the Covariant Quantization of Superstrings},
Class. Quant. Grav. 20 (2003) S395, hep-th/0302147\semi
P.A. Grassi, G. Policastro and
P. Van Nieuwenhuizen, 
{\it The Quantum Superstring as a WZNW Model},
Nucl. Phys. B676 (2004) 43, hep-th/0307056\semi 
P.A. Grassi and P. Van Nieuwenhuizen, 
{\it Gauging Cosets}, hep-th/0403209.}
\ref\kazama{Y. Aisaka and Y. Kazama, {\it A New First Class Algebra,
Homological Perturbation and Extension of Pure Spinor Formalism for
Superstring}, JHEP 0302 (2003) 017, hep-th/0212316\semi
Y. Aisaka and Y. Kazama, {\it Operator Mapping between RNS and
Extended Pure Spinor Formalisms for Superstring}, JHEP 0308 (2003) 047,
hep-th/0305221\semi 
Y. Aisaka and Y. Kazama, {\it Relating Green-Schwarz and Extended
Pure Spinor Formalisms by Similarity Transformation}, JHEP 0404 (2004)
070, hep-th/0404141.}\ref\chestone{M. Chesterman, {\it Ghost Constraints
and the Covariant Quantization of the Superparticle in Ten Dimensions},
JHEP 0402 (2004) 011, hep-th/0212261.}\ref\chesttwo{M. Chesterman,
{\it On the Cohomology and Inner Products of the Berkovits Superparticle
and Superstring}, hep-th/0404021.}.

In this paper, it will be shown how to perform functional integration
over the pure spinor ghosts\foot{Some features of this functional integration
method will appear in a separate paper 
with Sergei Cherkov \ref\cher{N. Berkovits
and S. Cherkov, in preparation.}.} 
by defining a Lorentz-invariant measure
and introducing appropriate ``picture-changing''
operators.\foot{Using the pure spinor version of the
superparticle, Chesterman has recently considered superparticle states
with non-standard boundary conditions for the pure spinor ghosts 
\chestone\chesttwo.
In independent work which appeared last month \chesttwo, 
Chesterman showed that these
states are related to standard superparticle states by an operator
$\psi_{-11}$ which plays the role of the 
picture-lowering operator described in this paper.

Also, in independent
work \ref\Grassi{P.A. Grassi, G. Policastro and
P. van Nieuwenhuizen, {\it Superstrings and WZNW Models},
hep-th/0402122.} which was announced after a seminar on
this paper, Grassi, Policastro and van Nieuwenhuizen 
used functional integration to define the measure factor in
their quantization approach without pure spinors. It would
be interesting to relate their functional integration
method with the method described here.}
These picture-changing operators will then be used to construct a
substitute for the $b$ ghost in a non-zero picture.
With these ingredients, it is straightforward to generalize
the tree amplitude prescription of \superp\
to a super-Poincar\'e covariant prescription for $N$-point 
$g$-loop amplitudes.
So there is no need to relax the pure spinor
constraint for the covariant computation of superstring amplitudes.

The need for picture-changing operators\foot{One 
can also use picture-changing operators
to construct a cubic pure spinor version of open superstring
field theory.
However, as in the cubic RNS version of open superstring field theory 
\ref\witcubic{E. Witten, {\it Interacting Field Theory of Open Superstrings},
Nucl. Phys. B276 (1986) 291\semi C.R. Preitschopf, C.B. Thorn and
S.A. Yost, {\it Superstring Field Theory}, Nucl. Phys. B337 (1990) 363\semi
I.Ya. Arefeva, A.S. Koshelev, D.M. Belov and P.B. Medvedev, 
{\it Tachyon Condensation in Cubic
Superstring Field Theory}, Nucl. Phys. B638 (2002) 3, hep-th/0011117.},
the action is expected to have gauge-invariance anomalies due
to picture-changing
operators at the string midpoint \ref\wendt{C. Wendt, {\it
Scattering Amplitudes and Contact Interactions in Witten's 
Superstring Field Theory}, Nucl. Phys. B314 (1989) 209\semi
I.Ya. Arefeva and P.B. Medvedev, {\it Anomalies in Witten's Field
Theory of the NSR String}, Phys. Lett. B212 (1988) 299\semi
N. Berkovits, {\it Review of Open Superstring Field Theory}, hep-th/0105230.}. 
It should be stressed that these
anomalies in cubic superstring field theory are caused by the use
of picture-changing operators in the presence of off-shell states
and do not imply
surface-term ambiguities in on-shell multiloop amplitudes.} 
in this formalism
is not surprising since, like
the bosonic $(\beta,\gamma)$ ghosts in the RNS formalism \fms, 
the pure
spinor ghosts are chiral bosons with worldsheet zero modes.
For $g$-loop amplitudes, the use of standard ``picture-zero'' vertex
operators implies that one needs to insert 11 ``picture-lowering'' operators
and $11g$ ``picture-raising'' operators to absorb the zero modes of the
11 pure spinor ghosts. As in the RNS formalism, the worldsheet derivatives
of these picture-changing operators are BRST trivial so, up to possible
surface terms, the amplitudes are independent of their locations on
the worldsheet. But unlike the RNS formalism, there is no need to
sum over spin structures, so there are no divergences at the boundary
of moduli space and surface terms can be safely ignored in the
loop amplitude computations.

Although the explicit computation of arbitrary loop amplitudes is 
complicated,
one can easily prove certain vanishing theorems by
counting zero modes of the fermionic superspace
variables. For example, S-duality of the Type IIB superstring implies
that $R^4$ terms in the low-energy effective action receive no 
perturbative corections above one-loop \ref\green{M.B. Green
and M. Gutperle, {\it Effects of D Instantons}, Nucl. Phys.
B498 (1997) 195, hep-th/9701093\semi M.B. Green and
P. Vanhove, {\it D-instantons, Strings and M-theory}, Phys. Lett.
B408 (1997) 122, hep-th/9704145\semi
Z. Bern, L. Dixon, D. Dunbar, M. Perelstein and J.S. Rozowsky,
{\it On the Relationship between Yang-Mills Theory and Gravity
and its Implication for Ultraviolet Divergences},
Nucl. Phys. B530 (1998) 401, hep-th/9802162.}. After much effort, this was
recently verified in the RNS formalism at two-loops \iengo\phong.
Using the formalism described here, this S-duality conjecture
can be easily verified for all loops. 

Similarly, one can easily prove
the non-renormalization theorem that 
massless $N$-point multiloop amplitudes vanish whenever $N<4$.
Assuming factorization,
this non-renormalization theorem implies the absence of divergences
near the boundary of moduli space \atick
\ref\martinec{E. Martinec, {\it Nonrenormalization Theorems and
Fermionic String Finiteness}, Phys. Lett. B171 (1986) 189.}.
The boundary of moduli space includes two types of degenerate surfaces:
surfaces where the radius $R$ of a handle shrinks to zero, and surfaces
which split into two worldsheets connected by a thin tube.
As explained in \atick, the first type of degenerate
surface does not lead to divergent amplitudes since, after including
the $\log(R)$ dependence coming from integration
over the loop momenta, the amplitude integrand diverges slower than $1/R$.
The second type of degenerate surface can lead to a divergent amplitude if
there is an onshell state propagating along the thin tube between the
two worldsheets. But when all external states are on one of the two
worldsheets, vanishing of the one-point function implies the absence
of this divergence. And when all but one of the external states are on
one of the two worldsheets, vanishing of the two-point function implies
the absence of this divergence. Finally, when there are at least two
external states on each of the two worldsheets, the divergence can be
removed by analytic continuation of the external momenta \atick. Note
that vanishing of the three-point function is not required for finiteness.

So if there are no unphysical
divergences in the interior of moduli space\foot
{In light-cone gauge, unphysical divergences in the interior
of moduli space could come from singularities between
colliding interaction points \greensite
\ref\mandfinite{S. Mandelstam,
{\it The n Loop String Amplitude: Explicit Formulas, Finiteness        
and Absence of Ambiguities},
Phys. Lett. B277 (1992) 82.}. 
In conformal gauge, there are no obvious potential
sources for these unphysical divergences in the interior of moduli space
since the amplitudes are
independent (up to surface terms)
of the locations of picture-changing operators.},
this non-renormalization theorem implies
that  
superstring multiloop amplitudes are perturbatively finite.
Previous attempts to prove this non-renormalization
theorem using the RNS formalism 
\ref\martinec{E. Martinec, {\it Nonrenormalization Theorems and
Fermionic String Finiteness}, Phys. Lett. B171 (1986) 189.} were
unsuccessful because they ignored
unphysical poles of the spacetime supersymmetry
currents \verlone\ and incorrectly assumed that the integrand
of the scattering amplitude was spacetime supersymmetric.
Using the GS formalism,
there are arguments for the non-renormalization theorem 
\ref\arguments
{A. Restuccia and J.G. Taylor, {\it Finiteness of Type II Superstring
Amplitudes}, Phys. Lett. B187 (1987) 267\semi R. Kallosh and A. Morozov,
{\it On Vanishing of Multiloop Contributions to 0,1,2,3 Point Functions
in Green-Schwarz Formalism for Heterotic String}, Phys. Lett. B207
(1988) 164.}, however, these arguments do not rule out the possibility of
unphysical divergences in the interior of moduli space from
contact term singularities between
light-cone interaction point operators \greensite.
Mandelstam was able to overcome this obstacle and prove
finiteness \mandfinite\
by combining different features of the RNS and GS formalisms.
However, the finiteness proof here is more direct than the proof of
\mandfinite\ since it is derived from a single formalism.

In section 2 of this paper, the super-Poincar\'e invariant 
pure spinor formalism
of \superp\ is reviewed. The first subsection reviews the worldsheet
action for the Green-Schwarz-Siegel matter variables and
the OPE's for the pure spinor ghosts.
The second subsection reviews the BRST operator and shows how physical
states are described by the BRST cohomology. The third subsection reviews
the computation of tree amplitudes using a measure factor determined by
cohomology arguments.

In section 3, functional integration over
the pure spinor ghosts is defined with the help of picture-changing 
operators. The first subsection shows how to define Lorentz-invariant
measure factors for integration over the pure spinor ghosts and their
conjugate momenta. The second subsection introduces picture-raising
and picture-lowering operators which are necessary for functional integration
over the bosonic ghosts. The third subsection shows that by inserting
picture-lowering operators, the tree amplitudes of section 2 can be 
computed using standard functional integration techniques.

In section 4, a composite $b$ ghost is defined
by requiring that the BRST variation of the $b$ ghost 
is a picture-raised version of the stress tensor.
The first subsection introduces a chain of operators of $+2$ conformal
weight which are useful for explicitly constructing the $b$ ghost.
The second subsection shows how the various terms in the composite
$b$ ghost can be expressed in terms of these operators.

In section 5, 
a super-Poincar\'e covariant prescription is given
for $N$-point $g$-loop amplitudes. The partition function for the
matter and ghost variables precisely cancel in this prescription, so one
only needs to compute correlation functions. The first and second
subsections show how to compute correlation functions for the matter
and ghost variables by separating off the zero modes and using the
free-field OPE's to functionally integrate over the non-zero modes.
The third subsection shows how to integrate over the zero modes using
the measure factor defined in section 3 for the pure spinor ghosts and
their conjugate momenta.

Finally, in section 6, the four-point
one-loop amplitude is computed and 
certain
vanishing theorems are proven using the multiloop prescription.
In the first subsection, the structure of Type II massless vertex
operators is reviewed. In the second subsection, the non-renormalization
theorem for less than four massless states is proven by zero-mode counting.
In the third subsection, the four-point massless one-loop amplitude is
explicitly computed up to an overall constant. And in the fourth subsection,
it is proven by zero-mode counting that
the $R^4$ term in the low-energy effective action does not receive
perturbative corrections above one loop.

\newsec{ Review of Super-Poincar\'e Covariant Pure Spinor Formalism}

\subsec{Worldsheet action}

The worldsheet variables in the Type IIB version of this formalism
include the Green-Schwarz-Siegel \ref\GScov{M.B. Green
and J.H. Schwarz, {\it Covariant Description of Superstrings},
Phys. Lett. B131 (1984) 367.}\ref\siegel{W. Siegel,
{\it Classical Superstring
Mechanics}, Nucl. Phys. B263 (1986) 93.} matter variables $(x^m,\t^\a,p_\a;
\bar\t^\a,\bar p_\a)$ for $m=0$ to 9 and $\a=1$ to 16, and the pure
spinor ghost variables $(\l^\a,w_\a;\bar\l^\a,\bar w_\a)$ where
$\l^\a$ and $\bar\l^\a$ are constrained to satisfy the pure spinor
conditions
\eqn\pure{\l^\a (\g^m)_{\a\b}\l^\b=0,\quad \bar\l^\a
(\gamma^m)_{\a\b}\bar\l^\b=0} 
for $m=0$ to 9. $(\gamma^m)_{\a\b}$ and $(\gamma^m)^{\a\b}$
are $16\times 16$ symmetric matrices which 
are the off-diagonal blocks
of the $32\times 32$ ten-dimensional $\Gamma$-matrices
and satisfy $(\g^{(m})_{\a\b} (\g^{n)})^{\b\g}=2\eta^{mn} \d_\a^\g$. 
For the 
Type IIA version
of the formalism,
the chirality of the
spinor indices on the right-moving variables is reversed, and
for the heterotic version, the right-moving variables are the same as
in the RNS formalism. 

In conformal gauge, the worldsheet action is
\eqn\action{S=\int d^2 z [-\half \p x^m\bar \p x_m -p_\a \pb\t^\a
-\bar p_\a \p\tb^\a + w_\a \pb\l^\a +\bar w_\a\p\lb^\a]}
where $\l^\a$ and $\lb^\a$ satisfy \pure.
The OPE's for the matter variables are easily computed to be
\eqn\wope{x^m(y) x^n(z) \to -\eta^{mn}\log |y-z|^2, \quad
p_\a(y)\t^\b (z) \to (y-z)^{-1} \d_\a^\b,}
however, the pure spinor constraint on $\l^\a$ prevents a direct
computation of its OPE's with $w_\a$.
As discussed in \superp,
one can solve the pure spinor constraint and express
$\l^\a$ in terms of eleven unconstrained free fields which manifestly
preserve a U(5) subgroup of the (Wick-rotated) Lorentz group. 
Although the OPE's of the unconstrained variables are not manifestly
Lorentz-covariant, 
all OPE computations involving $\l^\a$
can be expressed in a manifestly Lorentz-covariant
manner. So the non-covariant unconstrained
description of pure spinors is
useful only for verifying certain coefficients in the
Lorentz-covariant OPE's.

Because of the pure spinor constraint on $\l^\a$, the worldsheet
variables $w_\a$ contain the gauge invariance 
\eqn\gaugew{\d w_\a =\Lambda^m
(\g_m\l)_\a,}
so 5 of the 16 components of $w_\a$ can be gauged away. To preserve this
gauge invariance,
$w_\a$ can only appear in the gauge-invariant combinations 
\eqn\currents{N_{mn} = \half w_\a (\gamma_{mn})^\a{}_\b \l^\b,
\quad J= w_\a\l^\a ,}
which are the Lorentz currents and ghost current. As shown in 
\relating\ and
\cohom\ 
using either the U(5) 
or SO(8) unconstrained descriptions of pure spinors\foot{In reference \cohom\ 
for the
SO(8) description, the ghost current $J$ was not discussed. In terms of the
SO(8)-covariant variables of \cohom, 
$J = -2bc + s^a r^a -2\sum_{n=0}^\infty [n v^j_{(n)} w^j_{(n)}
+(n+\half) t^{\dot a}_{(n)} u^{\dot a}_{(n)}]$. Using the summation method
described in \cohom, it is straightforward to check that $J$ satisfies the 
OPE's described here.},
$N_{mn}$ and $J$ satisfy the Lorentz-covariant OPE's
\eqn\OPE{ N_{mn}(y) \l^\a(z) \to \half (y-z)^{-1} (\gamma_{mn}\l)^\a, \quad
J(y) \l^\a(z) \to (y-z)^{-1} \l^\a,}
$$N^{kl}(y) N^{mn}(z) \to 
- 3 (y-z)^{-2}
(\eta^{n[k} \eta^{l]m}) +
(y-z)^{-1}(\eta^{m[l} N^{k]n} -
\eta^{n[l} N^{k]m} ) 
,$$
$$ J(y) J(z) \to -4 (y-z)^{-2}, \quad J(y) N^{mn}(z) \to {\rm regular}, $$
$$N_{mn}(y) T(z) \to (y-z)^{-2} N_{mn}(z) ,\quad 
J(y) T(z) \to  -8(y-z)^{-3} + (y-z)^{-2} J(z),$$
where 
\eqn\stress{T=-\half \p x^m \p x_m - p_\a \p\t^\a + w_\a \p\l^\a}
is the left-moving stress tensor. From the OPE's of \OPE, one sees
that the pure spinor condition implies that the
levels for the Lorentz and ghost currents are $-3$ and $-4$,
and that the ghost-number anomaly is $-8$. 
Note that the total Lorentz current $M^{mn}=-\half (p\g^{mn} \t) + N^{mn}$
has level $k=4-3=1$, which coincides with the level of the RNS Lorentz
current $M^{mn}=\psi^m \psi^n$. The ghost-number anomaly of $-8$ 
will be related in section 3 to the pure spinor measure factor.

The stress tensor of \stress\
has no central charge since the $(+10-32)$ contribution
from the $(x^m,\t^\a,p_\a)$ variables is cancelled by the $+22$ contribution
from the eleven independent $(\l^\a,w_\a)$ variables. From its OPE's
with $N_{mn}$ and $J$, one learns that the stress tensor
can be expressed
in Sugawara form as\foot{There is a typo in the sign of the $\p J$ term
in references \relating\ and \ictp.} 
\eqn\sugawara{T =
-\half \p x^m \p x_m - p_\a \p\t^\a + {1\over{10}} :N^{mn} N_{mn}:
-{1\over 8} :J J: + \p J }
where the level $-3$ SO(9,1) current algebra contributes $-27$
to the central charge and the ghost current $J$ contributes $+49$.

\subsec{Physical states}

Physical open string states in this formalism are defined as
super-Poincar\'e covariant states of 
ghost-number $+1$ in the cohomology of the nilpotent BRST-like operator
\eqn\brst{Q = \oint \l^\a d_\a}
where 
\eqn\ddef{d_\a= p_\a -\half\g_{\a\b}^m \t^\b \p x_m -{1\over 8}
\g_{\a\b}^m \g_{m~\g\d}\t^\b\t^\g\p\t^\d}
is the supersymmetric Green-Schwarz constraint.
As shown by Siegel \siegel, $d_\a$ satisfies the OPE's 
\eqn\oped{d_\a(y) d_\b(z) \to -(y-z)^{-1} \g_{\a\b}^m \Pi_m,\quad 
d_\a(y) \Pi^m(z) \to  (y-z)^{-1} \g_{\a\b}^m \p\t^\b(z),}
$$d_\a(y) \p\t^\b(z) \to (y-z)^{-2} \d_\a^\b,\quad 
\Pi^m(y) \Pi^n(z) \to  -(y-z)^{-2} \eta^{mn},$$
where $\Pi^m = \p x^m +\half\t \g^m \p\t$
is the supersymmetric momentum and 
\eqn\defqq{
q_\a= \oint (p_\a +\half\g_{\a\b}^m \t^\b \p x_m +{1\over {24}}
\g_{\a\b}^m \g_{m~\g\d}\t^\b\t^\g\p\t^\d)}
is the supersymmetric generator satisfying
\eqn\sus{ \{q_\a,q_\b\}=
\g_{\a\b}^m\oint\p x_m, \quad [q_\a, \Pi^m(z)]=0,\quad
\{q_\a, d_\b(z)\}=0.}

To compute the massless spectrum of the open superstring\foot{Massless
vertex operators for the closed superstring will be reviewed in subsection
(6.1).},
note that
the most general vertex operator with zero conformal weight at zero momentum
and $+1$ ghost-number is 
\eqn\smax{V= \l^\a A_\a(x,\t),}
where $A_\a(x,\t)$ is a spinor superfield depending only on the worldsheet
zero modes of $x^m$ and $\t^\a$. Using the OPE that
$d_\a(y)~ f(x(z),\t(z))\to (y-z)^{-1} D_\a f$
where 
\eqn\susyd{D_\a=
{\p\over{\p\t^\a}} +\half \t^\b \g^m_{\a\b} \p_m }
is the supersymmetric
derivative, one can easily check that $QV=0$ and $\d V=Q\Lambda$
implies that $A_\a(x,\t)$ must satisfy $\l^\a\l^\b D_\a A_\b=0$
with the gauge invariance $\d A_\a=D_\a \Lambda$. But $\l^\a\l^\b
D_\a A_\b=0$ implies that 
\eqn\bianchi{D_\a A_\b + D_\b A_\a =\g^m_{\a\b} A_m}
for some vector superfield $A_m$ with the gauge transformations
\eqn\gaugeinv{\d A_\a = D_\a \Lambda, \quad \d A_m = \p_m\Lambda.}
In components, one can use \bianchi\ and \gaugeinv\ to gauge $A_\a$
and $A_m$ to the form
\eqn\opena{A_\a(x,\t)= e^{ik\cdot x}(\half a_m(\g^m\t)_\a -{1\over 3}
(\xi\g_m\t)(\g^m\t)_\a
 + ... ),}
$$A_m(x,\t) = e^{ik\cdot x}(a_m + (\xi\g^m\t) + ...),$$
where $k^2 = k^m a_m = k^m (\g_m\xi)_\a =0,$ and
$...$ involves products of $k_m$ with $a_m$ or $\xi^\a$.
So \bianchi\ and \gaugeinv\ are the equations of motion and 
gauge invariances of the ten-dimensional super-Maxwell multiplet, 
and the cohomology at ghost-number $+1$ of $Q$ correctly describes the
massless spectrum of the open superstring \ref\howe{P. Howe,
{\it Pure Spinor Lines in Superspace and Ten-Dimensional Supersymmetric
Theories}, Phys. Lett. B258 (1991) 141.}. 

To compute the massive spectrum, one needs to consider the cohomology
of vertex operators which have non-zero conformal weight at zero
momentum. This was done with Chandia in \ref\massive
{N. Berkovits and O. Chand\'{\i}a, {\it Massive Superstring Vertex Operator in
D=10 Superspace}, JHEP
0208 (2002) 040, hep-th/0204121.} for the massive spin-two
multiplet and gave for the first time its equations of motion in 
ten-dimensional superspace. To prove that the cohomology of $Q$
reproduces the superstring spectrum at arbitrary mass level, the
SO(8)-covariant description was used in \cohom\ to
solve the pure spinor constraint, and the resulting BRST cohomology
was shown to be equivalent to the light-cone GS spectrum.

In addition to describing the spacetime fields at ghost-number $+1$,
the cohomology of $Q$ can also be used to describe the spacetime ghosts
at ghost-number zero, the spacetime antifields at ghost-number $+2$,
and the spacetime antighosts at ghost-number $+3$ \ref\siegf{W. Siegel,
{\it First Quantization and Supersymmetric Field Theories}, 
Stony Brook 1991 Strings and Symmetries Proceedings (1991).}\ref\superpart
{N. Berkovits, {\it Covariant Quantization of the Superparticle using
Pure Spinors}, JHEP 09 (2001) 016, hep-th/0105050.}.
For example, the
super-Yang-Mills ghost at ghost-number zero
is described by 
the vertex operator $V=\Lambda$, the super-Yang-Mills antifields
at ghost-number two
are described by the vertex operator $V=\l^\a\l^\b A^*_{\a\b}(x,\t)$,
and the Yang-Mills antighost at ghost-number three
is described by the vertex operator
$V=(\l\g^m\t)(\l\g^n\t)(\l\g^p\t)(\t\g_{mnp}\t)$.
As was shown in \superpart, the conditions $QV=0$ and $\d V=Q\Lambda$
imply the correct equations of motion and gauge invariances for these
ghosts, antifields and antighosts. 

\subsec{Tree-level prescription}

As in the bosonic string, the prescription for $N$-point open
string tree amplitudes
in this formalism requires three dimension-zero
vertex operators $V$ and $N-3$ dimension-one vertex operators $U$
which are integrated over the real line. Normally, one defines the
dimension-one vertex operators by
$U(z)=\{\oint b, V(z)\}$ where $b(z)$ is the dimension-two field 
satisfying $\{Q,b(z)\}=T(z)$. Since $QV=0$ and $[\oint T, V(z)]=\p V(z)$,
this relation implies that $QU=\p V$. 

In this formalism of the superstring, there are no states of negative
ghost number since the variable $w_\a$ can only appear through
the ghost-number zero operators $N_{mn}$ and $J$. So one cannot
construct a $b$ ghost satisfying $\{Q,b\}=T$. Nevertheless, since there
is no BRST cohomology for unintegrated dimension-one operators, one
is guaranteed 
that $QV=0$ implies that $\p V$ can be written as $QU$ for some $U$. 
For example, for the 
super-Maxwell vertex
operator $V=\l^\a A_\a$, one can check that 
\eqn\supermax{U = \p \t^\a A_\a (x,\t) + 
\Pi^m A_m (x,\t) + d_\a W^\a (x,\t) +\half N^{mn} {\cal F}_{mn}(x,\t)}
satisfies $QU= \p(\l^\a A_\a)$
where $A_m = {1\over 8}D_\a \g_m^{\a\b} A_\b$ is the vector gauge superfield,
$W^\b = {1\over{10}}\g_m^{\a\b} (  D_\a A^m -\p^m A_\a)$ is the
spinor superfield strength, and ${\cal F}_{mn} =
{1\over 8}D_\a (\g_{mn})^\a{}_\b W^\b = \p_{[m} A_{n]}$ is the
vector superfield strength.

In reference \superp, open string tree amplitudes were defined by
the correlation function
\eqn\ampl{{\cal A}
 =  \langle ~V_1(z_1) ~ V_2(z_2) ~
V_3(z_3) ~\int dz_4 U_4(z_4) ... \int dz_N U_N(z_N) ~\rangle.}
To compute this correlation function, the OPE's of \OPE\ and \oped\ were 
used to perform the functional integration
over the non-zero modes of the worldsheet variables.
Since $N_{mn}$, $J$ and $d_\a$ are fields of $+1$
conformal weight with no zero modes on a sphere, the dependence of 
the correlation function on their locations is completely determined by
the singularities in their OPE's. For example, the OPE's of $d_\a(z)$ imply
that\foot{To keep supersymmetry manifest, it is convenient to use the
OPE's of \oped\ for $d_\a$ instead of using the free-field OPE's of
\wope\ for $p_\a$.}
\eqn\corex{\langle
d_\a(z) \Pi^m(u) \p\t^\b(v) d_\g(w) A_\d(x(y),\t(y))\rangle =}
$$
\g^m_{\a\rho} (z-u)^{-1} \langle\p\t^\rho(u)
\p\t^\b(v) d_\g(w) A_\d(x(y),\t(y))\rangle$$
$$+\d_\a^\b
(z-v)^{-2} \langle\Pi^m(u) d_\g(w) A_\d(x(y),\t(y))\rangle$$
$$+\g_{\a\g}^n (z-w)^{-1}
\langle\Pi^m(u)\p\t^\b(v) \Pi^n(w) A_\d(x(y),\t(y))\rangle$$
$$+ (z-y)^{-1}
\langle\Pi^m(u)\p\t^\b(v) d_\g(w) D_\a A_\d(x(y),\t(y))\rangle.$$
And the OPE's of $N_{mn}(z)$ imply that
\eqn\coren{\langle N_{mn}(z) N_{pq}(u) \l^\a(v)\l^\b(w)\l^\g(y)\rangle = }
$$=(z-u)^{-1} 
\langle (\eta_{p[n} N_{m]q}(u) -\eta_{q[n} N_{m]p}(u)) \l^\a(v)\l^\b(w)
\l^\g(y)\rangle $$
$$
-3(z-u)^{-2} \eta_{q[m}\eta_{n]p} \langle \l^\a(v)\l^\b(w)\l^\g(y)\rangle$$
$$
+\half (z-v)^{-1} \langle N_{pq}(u) (\g_{mn}\l(v))^\a\l^\b(w)\l^\g(y)\rangle$$
$$
+\half (z-w)^{-1} \langle N_{pq}(u) \l^\a(v)(\g_{mn}\l(w))^\b\l^\g(y)\rangle$$
$$
+\half (z-y)^{-1} \langle N_{pq}(u) \l^\a(v)\l^\b(w)(\g_{mn}\l(y))^\g\rangle.$$

After using their OPE's to remove all $N_{mn}$'s, $J$'s and $d_\a$'s from
the correlation function, one can replace all remaining $\l^\a$ and
$\t^\a$ variables by their zero modes.
But since it was not known
how to perform the functional integration over the remaining zero modes
of the worldsheet scalars $\l^\a$ and $\t^\a$, an ansatz had to
be used for deciding which zero modes of $\l^\a$ and $\t^\a$
need to be present for non-vanishing amplitudes.

For tree amplitudes in bosonic string theory, the zero-mode prescription
coming from functional integration is 
\eqn\zerom{\langle~c~\p c~
\p^2 c~\rangle=1}
where $c$ is the worldsheet ghost of dimension $-1$. Since $c\p c\p^2 c$
is the vertex operator of $+3$ ghost-number for the Yang-Mills antighost
\siegf,
it is natural to use the ansatz that non-vanishing
correlation functions in this formalism must also be proportional to 
the vertex operator for the Yang-Mills antighost. As discussed in
the previous subsection, this vertex operator
in the pure spinor formalism is 
\eqn\antigh{V=(\l\g^m\t)(\l\g^n\t)(\l\g^p\t)(\t\g_{mnp}\t),}
which is the unique state in the BRST cohomology at $+3$ ghost-number.
So the zero mode prescription for tree amplitudes in the pure spinor
formalism is 
\eqn\zeromodep{
\langle(\l\g^m\t)(\l\g^n\t)(\l\g^p\t)(\t\g_{mnp}\t)\rangle =1.}

For later use, it will be convenient to write \antigh\ as
\eqn\conven{V= 
{\cal T}_{((\a_1\a_2\a_3))[\d_1\d_2\d_3\d_4\d_5]} \l^{\a_1}\l^{\a_2}
\l^{\a_3}
\t^{\d_1}\t^{\d_2}
\t^{\d_3}\t^{\d_4}
\t^{\d_5},}
where
\eqn\Tdef{{\cal T}_{((\a_1\a_2\a_3))[\d_1\d_2\d_3\d_4\d_5]}}
is a constant Lorentz-invariant tensor and
the notation $((\a_1\a_2\a_3))[\d_1\d_2\d_3\d_4\d_5]$
signifies that the tensor is
symmetric and $\g$-matrix traceless (i.e.
$\g_m^{\a_1\a_2} {\cal T}_{((\a_1\a_2\a_3))[\d_1 ...\d_5]}=0$)
in the
first three indices, and antisymmetric in the last five indices. 
This tensor is uniquely defined up to rescaling and can be computed
by starting with
$\g^m_{\a_1\d_1}\g^n_{\a_2\d_2}\g^p_{\a_3\d_3}(\g_{mnp})_{\d_4\d_5},$
then symmetrizing in the $\a$ indices, antisymmetrizing in the $\d$ indices,
and subtracting off the $\g$-matrix trace in the $\a$ indices.
Similarly, one can define the tensor
\eqn\Tinvdef{({\cal T}^{-1})^{((\a_1\a_2\a_3))[\d_1\d_2\d_3\d_4\d_5]}}
by starting with
$(\g^m)^{\a_1\d_1}(\g^n)^{\a_2\d_2}(\g^p)^{\a_3\d_3}(\g_{mnp})^{\d_4\d_5}$
and following the same procedure.

Using the properties of spinors in ten dimensions, it will be possible to
prove various identities satisfied by 
${\cal T}_{((\a_1\a_2\a_3))[\d_1\d_2\d_3\d_4\d_5]}.$
For example, there are no Lorentz scalars which can be constructed out
of four $\l$'s and four $\t$'s, which implies that 
\eqn\identone{
\d_{((\a_1}^{\d_5} {\cal T}_{\a_2\a_3\a_4))[\d_1\d_2\d_3\d_4\d_5]} =0}
and that 
\eqn\identtwo{{\cal T}_{((\a_1\a_2\a_3))[\d_1\d_2\d_3\d_4\d_5]}\neq
{\cal S}_{((\a_1\a_2\a_3\a_4))[\d_1\d_2\d_3\d_4} \d^{\a_4}_{\d_5]}}
for any tensor ${\cal S}_{((\a_1\a_2\a_3\a_4))[\d_1\d_2\d_3\d_4]}$. 
Furthermore,
there are no Lorentz scalars which can be constructed out
of two $\l$'s and six $\t$'s, which implies that 
\eqn\identthree{
{\cal T}_{((\a_1\a_2\a_3))[\d_1\d_2\d_3\d_4\d_5} \d^{\a_3}_{\d_6]} =0}
and that
\eqn\identfour{{\cal T}_{((\a_1\a_2\a_3))[\d_1\d_2\d_3\d_4\d_5]}\neq
\d_{((\a_1}^{\d_6}{\cal S}_{\a_2\a_3))[\d_1\d_2\d_3\d_4\d_5\d_6]} }
for any tensor ${\cal S}_{((\a_2\a_3))[\d_1\d_2\d_3\d_4\d_5\d_6]}$. 
Finally, one can check that 
$$(\l\g^q\t)(\l\g^m\t)(\l\g^n\t)(\l\g^p\t)(\t\g_{mnp}\t)=0$$
for
any $q$, which implies that 
\eqn\identfive{
\d_{((\a_1}^{\k} {\cal T}_{\a_2\a_3\a_4))[\d_1\d_2\d_3\d_4\d_5}\g^q_{\d_6]\k}
 =0.}

Using \zeromodep,
the zero-mode prescription for tree amplitudes is 
$$\langle {\cal T}_{((\a_1\a_2\a_3))[\d_1 ...
\d_5]}\l^{\a_1}\l^{\a_2}\l^{\a_3}
\t^{\d_1} ...
\t^{\d_5} \rangle =1.$$ 
In other words, suppose that ${\cal A}=\langle \l^\a\l^\b\l^\g
f_{\a\b\g}(\t) \rangle$ is the expression
one gets
after integrating out the non-zero modes, where $f_{\a\b\g}$ is
some complicated function of the polarizations and momenta of the
external states. Then the scattering amplitude is
defined as
\eqn\zpres{{\cal A}=({\cal T}^{-1})^{((\a\b\g))[\d_1 ...\d_5]} 
{\p\over{\p\t^{\d_1}}} 
...
{\p\over{\p\t^{\d_5}}} f_{\a\b\g}(\t) .}
Using this prescription and the identities of
\identone -\identfive, it was shown in \superp\ that on-shell
tree amplitudes are gauge-invariant and supersymmetric. And it was
shown in \vallilo\ 
and \relating\ that this tree amplitude prescription agrees with
the standard RNS prescription. However, it was unclear how to generalize
this prescription to loop amplitudes since it was not derived from
functional integration. In the next section, it will be shown how
to use picture-changing operators to resolve this problem.

\newsec{Functional Integration and Picture-Changing Operators}

As reviewed in section (2.1), the gauge invariance of \gaugew\
implies that pure spinor ghosts can only appear
through the operators 
$\l^\a$, $N_{mn}$ and $J$.
Correlation functions for the non-zero modes of these operators are easily
computed using the OPE's of \OPE. However, after integrating out the non-zero
worldsheet modes, one still has to functionally integrate over the worldsheet
zero modes. Because $\l^\a$ has zero conformal weight and satisfies
the pure spinor constraint
\eqn\puresc{\l\g^m\l=0,}
$\l^\a$ has 11 independent zero modes
on a genus $g$ surface. And because $N_{mn}$ and $J$ have $+1$ conformal
weight and are defined from gauge-invariant combinations of $w_\a$,
they have $11g$ independent zero modes on a genus $g$ surface. Note that
\puresc\ 
implies that $N_{mn}=\half (w\g_{mn}\l)$ and $J=w\l$ are related by
the equation\massive
\eqn\Ncons{:N^{mn} \l^\a: \g_{m\a\b} -\half :J\l^\a: \g^n_{\a\b}
= 2\g^n_{\a\b}\p\l^\a}
where the normal-ordered product is defined by 
$:U^A(z)\l^\a(z): = \oint dy (y-z)^{-1} U^A(y)\l^\a(z).$ (The coefficient
of the $\p\l^\a$ term is determined by computing the double pole of the 
left-hand side of \Ncons\ with $J$.) Just as \puresc\
implies that all 16 components of $\l^\a$ can be expressed in terms
of 11 components, equation \Ncons\ implies that all
45 components of
$N^{mn}$ can be expressed in terms of $J$ and ten components of
$N^{mn}$.

Because of the constraints of \puresc\ and \Ncons, it is not immediately
obvious how to functionally integrate over the pure spinor ghosts. However,
as will be shown in the following subsection, there is a natural
Lorentz-invariant measure factor for the pure spinor ghosts which can be
used to define functional integration.

\subsec{Measure factor for pure spinor ghosts}

A Lorentz-invariant measure factor for the $\l^\a$ zero modes can
be obtained by noting that 
\eqn\measured{(d^{11}\l)^{[\a_1\a_2 ...\a_{11}]}\equiv d\l^{\a_1} \wedge
d\l^{\a_2} \wedge ... \wedge d\l^{\a_{11}}}
satisfies the identity
\eqn\idend{ \l^\b \g^m_{\a_1 \b} 
(d^{11}\l)^{[\a_1 \a_2 ...\a_{11}]}= 0}
because $\l\g^m d\l=0$.
Using the properties of pure spinors, this implies that all 
${{16!}\over{5! 11!}}$ components
of 
$(d^{11}\l)^{[\a_1 ...\a_{11}]}$ 
are related to each other by a Lorentz-invariant
measure factor $[{\cal D}\l]$ of $+8$ ghost number which is defined by 
\eqn\mld{ 
(d^{11}\l)^{[\a_1 ...\a_{11}]}= 
[{\cal D}\l ]~~
(\e {\cal T})_{((\b_1\b_2\b_3))}^{[\a_1 ... \a_{11}]} \l^{\b_1}\l^{\b_2}
\l^{\b_3} }
where
$$(\e {\cal T})_{((\b_1\b_2\b_3))}^{[\a_1 ... \a_{11}]} =
\e^{\a_1 ... \a_{16}}
{\cal T}_{((\b_1\b_2\b_3))[\a_{12} ... \a_{16}]}$$
and \idend\ is implied by \mld\ using the identity of \identfive.
In other words, for any choice of $[\a_1 ... \a_{11}]$, one can define
the Lorentz-invariant measure $[{\cal D}\l]$ by the formula
\eqn\mldother{ 
[{\cal D}\l ] = 
(d^{11}\l)^{[\a_1 ...\a_{11}]} ~ 
[(\e {\cal T})_{((\b_1\b_2\b_3))}^{[\a_1 ... \a_{11}]} \l^{\b_1}\l^{\b_2}
\l^{\b_3}]^{-1}, }
where there is no sum over $[\a_1 ...\a_{11}]$ in \mldother.

One can similarly construct a Lorentz-invariant measure factor for
the $N^{mn}$ and $J$ zero modes from
\eqn\measureN{(d^{11}N)^{[[m_1 n_1][m_2 n_2]...[m_{10}n_{10}]]}\equiv
dN^{[m_1 n_1]} \wedge
dN^{[m_2 n_2]} \wedge ... \wedge dN^{[m_{10} n_{10}]} \wedge dJ.}
Using the constraint of \Ncons\ and keeping $\l^\a$ fixed while
varying $N^{mn}$ and $J$, one finds that
\measureN\ satisfies the identity
\eqn\idenN{ (\l\g_{m_1})_\a 
(d^{11}N)^{[[m_1 n_1][m_2 n_2]...[m_{10}n_{10}]]}=0.}
Using the properties of pure spinors, this implies that all
${{45!}\over{10! 35!}}$ components of
$$(d^{11}N)^{[[m_1 n_1][m_2 n_2]...[m_{10}n_{10}]]}$$
are related to each
other by a Lorentz-invariant measure factor $[{\cal D}N]$
of $-8$ ghost number
which is defined by
\eqn\mlN{ 
(d^{11}N)^{[[m_1 n_1][m_2 n_2]...[m_{10}n_{10}]]} =
[{\cal D}N] }
$$\left( (\l\g^{m_1 n_1 m_2 m_3 m_4}\l)
(\l\g^{m_5 n_5 n_2 m_6 m_7}\l)
(\l\g^{m_8 n_8 n_3 n_6 m_9}\l)
(\l\g^{m_{10} n_{10} n_4 n_7 n_9}\l) + {\rm permutations}\right)$$
where the permutations are antisymmetric under the exchange of 
$m_j$ with $n_j$, and also antisymmetric under the exchange of 
$[m_j n_j]$ with $[m_k n_k]$.
Note that the index structure on the right-hand side
of \mlN\ has been chosen such the 
expression is non-vanishing after summing over the permutations.

After using the OPE's of \OPE\ to integrate out the non-zero modes of the
pure spinor ghosts on a genus $g$ surface, one will obtain an expression
\eqn\obt{{\cal A}=\langle f(\l, N_1,J_1,N_2, J_2, ..., N_g, J_g)\rangle}
which only depends on the 11 worldsheet zero modes of $\l$, and on
the 11$g$ worldsheet zero modes of $N$ and $J$. 
Using the Lorentz-invariant
measure factors defined in \mld\ and \mlN, the natural definition for
functional integration over these zero modes is
\eqn\func{{\cal A}=\int [{\cal D}\l]
[{\cal D}N_1]
[{\cal D}N_2]
...
[{\cal D}N_g]
f(\l, N_1,J_1,N_2, J_2, ..., N_g, J_g).}
Note that with this definition, 
$f(\l, N_1,J_1,N_2, J_2, ..., N_g, J_g)$ must carry ghost number 
$-8+8g$ to give a non-vanishing functional integral, which agrees with
the $-8$ ghost-number anomaly in the OPE of $J$ with $T$.
It will now be shown how the functional integral of \func\ can be explicitly
computed with the help of picture-changing operators.

\subsec{Picture-changing operators}

As is well-known from the work of Friedan-Martinec-Shenker \fms\ and
Verlinde-Verlinde \verlone\verltwo, picture-changing
operators are necessary in the RNS formalism because of the bosonic
$(\b,\g)$ ghosts. 
Since the picture-raising and picture-lowering operators involve the
delta functions $\d(\b)$ and $\d(\g)$, insertion of these operators
in loop amplitudes are needed to absorb
the zero modes of the $(\b,\g)$
ghosts on a genus $g$ surface.\foot{ 
In the RNS formalism, it is convenient to bosonize the $(\b,\g)$ ghosts
as $\b=\p\xi e^{-\phi}$ and $\gamma = \eta e^\phi$
since the spacetime supersymmetry generator involves a spin field constructed
for the chiral boson $\phi$. The delta functions $\d(\b)$ and $\d(\g)$
can then be expressed in terms of $\phi$ as $\d(\b)= e^\phi$ and
$\d(\g)= e^{-\phi}$. However, in the pure spinor formalism, there is
no advantage to performing such a bosonization since all operators can be
expressed directly in terms of $\l^\a$, $N^{mn}$ and $J$. Since
functional integration over the $\phi$ chiral boson can give rise
to unphysical poles in the correlation functions, the fact that all
operators in the pure spinor formalism can be expressed in terms of
$(\l^\a,N^{mn},J)$ implies that there are no unphysical poles in pure
spinor correlation functions.}
Up to possible surface terms, the amplitudes are independent
of the worldsheet positions of these operators since the 
worldsheet derivatives of the picture-changing operators are BRST-trivial. 
The surface terms come from pulling the BRST operator through the $b$
ghosts to give total derivatives in the worldsheet moduli. If the
correlation function diverges near the boundary of moduli space, these
surface terms can give finite contributions which need to be treated
carefully. 
As will now be shown, functional integration over the bosonic 
ghosts in the pure spinor formalism also requires picture-changing
operators with similar properties to those of the RNS formalism. 
However, since the correlation functions in this formalism do
not diverge near the boundary of moduli space, there are no
subtleties due to surface terms. 

To absorb the zero modes of $\l^\a$, $N_{mn}$ and $J$, picture-changing
operators in the pure spinor formalism will involve the delta-functions
$\d(C_\a \l^\a)$, $\d(B_{mn} N^{mn})$ and $\d(J)$ where
$C_\a$ and $B_{mn}$ are constant spinors and antisymmetric tensors.
Although these constant spinors and tensors are needed for the construction
of picture-changing operators,
it will be shown that scattering amplitudes are independent of
the choice of $C_\a$ and $B_{mn}$, so Lorentz invariance is preserved.
As will be discussed later, this Lorentz invariance can be made manifest
by integrating over all choices of $C_{\a}$ and $B_{mn}$.
Note that the use of constant spinors and tensors in picture-changing
operators is unrelated 
to the pure spinor constraint, and is necessary whenever the bosonic ghosts
are not Lorentz scalars.

As in the RNS formalism, the picture-changing
operators will be BRST-invariant with
the property that their worldsheet derivative is BRST-trivial.
A ``picture-lowering'' operator $Y_C$ with these properties is
\eqn\plo{Y_C = C_\a \t^\a \d(C_\b \l^\b)}
where $C_\a$ is any constant spinor.
Note that $Q Y_C = (C_\a\l^\a) \d(C_\b\l^\b)=0$ and
\eqn\secy{\p Y_C = (C \p\t)\d(C\l) + (C\t)(C\p\l)\p\d(C\l)
= Q[ (C\p\t)(C\t) \p\d(C\l)]}
where $\p\d(x) \equiv {\p\over\p x}\d(x)$ is defined using the
usual rules for derivatives of delta functions, e.g. 
$x \p\d(x) = -\d(x)$.\foot{Throughout this paper, the symbol $\p$
will denote the worldsheet derivative ${\p\over{\p z}}$ except when
$\p$ acts on a delta function. When acting on a delta function,
$\p\d(x)$ will denote ${\p\over{\p x}} \d(x)$.}

Although $Y_C$ is not spacetime-supersymmetric,
its supersymmetry variation is BRST-trivial since
\eqn\susyy{q_\a Y_C = C_\a\d(C\l) = - C_\a (C\l) \p\d(C\l)=
Q[ -C_\a (C\t) \p\d(C\l)].}
Similarly, $Y_C$ is not Lorentz invariant, but its Lorentz variation
is BRST-trivial since 
\eqn\lorentzy{M^{mn} Y_C = \half (C\g^{mn}\t)\d(C\l) +\half (C\t)
(C\g^{mn}\l)\p\d(C\l) =
Q[ \half (C\g^{mn}\t)(C\t)\p\d(C\l)].}
So different choices of $C_\a$ only change $Y_C$ by a 
BRST-trivial quantity,
and any on-shell amplitude computations involving insertions of $Y_C$
will be Lorentz invariant and spacetime supersymmetric up to possible
surface terms. 
The fact that Lorentz invariance is preserved only up
to surface terms is unrelated to the pure spinor constraint, and is
caused by the bosonic ghosts not being Lorentz scalars.

One can also construct BRST-invariant operators
involving $\d (B^{mn} N_{mn})$ and $\d (J)$
with the property that their worldsheet derivative is BRST-trivial.
These ``picture-raising'' operators will be called $Z_B$ and $Z_J$
and are defined
by
\eqn\pro{Z_B = \half B_{mn} (\l\g^{mn}d) \d(B^{pq}N_{pq}),\quad Z_J=
(\l^\a d_\a) \d(J),}
where $B_{mn}$ is a constant antisymmetric tensor.
To eliminate the need for normal-ordering in $Z_B$, it will be convenient
to choose $B_{mn}$ such that it satisfies
\eqn\Bconv{ B_{mn} B_{pq} (\g^{mn}\g^p)_\a{}^\b =0.}
(To give a concrete example, 
$B_{mn}$ satisfies \Bconv\ if its only non-zero components
are in the directions $B_{13}=iB_{23}= -B_{24} =iB_{14}.$)
With this choice, $(\l\g^{mn}d)B_{mn}$ has no pole with $B^{pq} N_{pq}$,
and therefore $(\l\g^{mn}d)B_{mn}$ has no pole with $\d(B^{pq}N_{pq})$.

Since $\l^\a d_\a$ has a pole with $\d(J)$, it naively appears that
$Z_J$ needs to be regularized. However, $\l^\a d_\a$ is the BRST current
which has no poles anywhere else on the surface. Since $Z_J$ will only
be needed on surfaces of non-zero genus, and since any function with a 
single pole on such surfaces must be a constant function, $\l^\a d_\a$
has no pole with $\d(J)$ and therefore $Z_J$ does not need to be regularized.

$Z_B$ and $Z_J$ satisfy the properties of picture-changing operators since
\eqn\propone{QZ_B = -{1\over 4} B_{mn} B_{pq}(\l\g^{mn}d)(\l\g^{pq}d)
\p\d(B N) - \half B_{mn}\Pi_p (\l\g^{mn}\g^p \l) \d(BN) =0,}
$$QZ_J = (\l_\a d^\a) (\l_\b d^\b) \p\d(J) - \Pi_m (\l\g^m \l)\d(J)=0,$$
$$\p Z_B = \half B_{mn} \p(\l\g^{mn}d) \d(BN)+\half B_{mn}(\l\g^{mn}d) 
B_{pq}\p N^{pq} \p\d(BN) = Q[ B_{pq} \p N^{pq} \d(BN)], $$
$$\p Z_J = \p(\l d)\d(J) + (\l d)\p J \p\d(J) = Q[-\p J\d(J)].$$
Furthermore, $Z_B$ and $Z_J$ are manifestly spacetime supersymmetric and
the Lorentz transformation of $Z_B$ is BRST-trivial since
\eqn\mzb{M^{mn} Z_B = 
\eta^{p[m}\d^{n]}_r [ B_{pq} (\l\g^{qr}d) \d(BN)
+ B_{st} (\l\g^{st}d) B_{pq} N^{qr} \p\d(BN)]}
$$
= Q[2 \eta^{p[m}\d^{n]}_r B_{pq} N^{qr} \d(BN)].$$
So different choices of $B_{mn}$ only change $Z_B$ by a
BRST-trivial quantity.

With these picture-changing operators, the pure spinor measure factors
of subsection (3.1)
can be used to compute arbitrary loop amplitudes with functional
integration methods. But before discussing loop amplitudes, it will be
useful to show how these functional integration methods reproduce the
tree amplitude prescription of subsection (2.3).

\subsec{Functional integration computation of tree amplitude}

For tree amplitudes, $\l^\a$ has eleven zero modes so one needs
to insert eleven picture-lowering operators $Y_{C_1}(y_1)...Y_{C_{11}}(y_{11})$
into the correlation function where the choices of $C_I$ and $y_I$
for $I=1$ to 11 are arbitrary. It will now be shown that
functional integration with these insertions reproduces the correct tree
amplitude prescription.
Using the notation of subsection (2.3),
the $N$-point open string tree amplitude is computed by the 
correlation function
\eqn\ampltwo{{\cal A}
 =  \langle ~V_1(z_1) ~ V_2(z_2) ~
V_3(z_3) ~\int dz_4 U_4(z_4) ... \int dz_N U_N(z_N) ~
Y_{C_1}(y_1)...Y_{C_{11}}(y_{11}) \rangle}
where $V$ and $U$ are the unintegrated and integrated vertex operators
and $Y_{C_I}(y_I) = C_{I\a} \t^\a(y_I) \d(C_I\l(y_I))$. 

To compare with the prescription of \ampl, it is convenient to fix
$(z_1,z_2,z_3)$ at finite points on the worldsheet and to insert all
eleven picture-lowering operators at $y_I=\infty$. With this
choice, there are no
contributions from the OPE's of the picture-lowering
operators with the $N$ vertex 
operators. Also, there are no singular OPE's between the picture-lowering
operators since $\d(C_1\l)$ has a pole with $\d(C_2\l)$ only
when $C_{1\a}$ is proportional to $C_{2\a}$,
which implies that $(C_1\t)$ has a zero with $(C_2\t)$.
After integrating over the non-zero modes of the worldsheet
fields, one is left with
the expression
\eqn\ampleft{
{\cal A}
 = \langle ~ \l^\a\l^\b \l^\g f_{\a\b\g}(\t) 
(C_1\t) ...(C_{11}\t) \d(C_1\l) ...\d(C_{11}\l)
~\rangle}
where $f_{\a\b\g}(\t)$ is the same function as in the computation of \zpres.
To integrate over the $\t^\a$ and $\l^\a$ zero modes, use the standard
$\int d^{16}\t$ measure factor and the pure spinor measure factor of \mld\ 
to obtain
\eqn\amplefttwo{
{\cal A}
 = \int d^{16}\t \int [{\cal D}\l] ~ \l^\a\l^\b\l^\g f_{\a\b\g}(\t) 
(C_1\t) ...(C_{11}\t) \d(C_1\l) ...\d(C_{11}\l) }
$$= \int d^{16}\t (\e{\cal T}^{-1})^{((\a\b\g))}_{[\rho_1 ...\rho_{11}]}
\int d\l^{\rho_1} ... d\l^{\rho_{11}} f_{\a\b\g}(\t)
(C_1\t)...(C_{11}\t) \d(C_1\l) ...\d(C_{11}\l). $$

In general, 
\eqn\defFF{\int
d\l^{\rho_1}... d\l^{\rho_{11}} \d(C_1\l) ...\d(C_{11}\l)}
is a complicated
function of $C_I$ because of the Jacobian coming from expressing
$\l^\rho$ in terms of $(C_I\l)$. However, since the Lorentz variation
of $Y_C$ is BRST-trivial, the amplitude is independent (up to possible
surface terms) of the choice of $C_I$. This implies that if one
integrates \amplefttwo\ over all possible choices for $C_I$ with
a measure factor $[{\cal D} C]$ satisfying
$\int [{\cal D}C] =1$, the amplitude is unchanged. Note that \amplefttwo\
is manifestly invariant under rescalings of $C_I$, so $C_I$ can
be interpreted as a projective coordinate. 

So one can express the amplitude in Lorentz-covariant form as 
\eqn\aimpl{
{\cal A} = 
(\e{\cal T}^{-1})^{((\a\b\g))}_{[\rho_1 ...\rho_{11}]}
\int d^{16}\t~ 
\t^{\k_1} ... \t^{\k_{11}} f_{\a\b\g}(\t)}
$$
\int[{\cal D}C] \int d\l^{\rho_1} ...d\l^{\rho_{11}}
C_{1\k_1} ... C_{11\k_{11}}  \d(C_1\l) ...\d(C_{11}\l).$$
By Lorentz invariance,
\eqn\lorentzi{\int[{\cal D}C] \int d\l^{\rho_1} ...d\l^{\rho_{11}}
C_{1\k_1} ... C_{11\k_{11}}  \d(C_1\l) ...\d(C_{11}\l)
=c \d^{[\rho_1}_{\k_1} ... \d^{\rho_{11}]}_{\k_{11}}}
where $c$ is a normalization factor which is determined from 
\eqn\detc{
\int[{\cal D}C] \int (C_{1\rho_1} d\l^{\rho_1}) ...
(C_{11\rho_{11}} d\l^{\rho_{11}})
\d(C_1\l)...
\d(C_{11}\l) =1.}
So 
\eqn\aimpf{
{\cal A} = c 
(\e{\cal T}^{-1})^{((\a\b\g))}_{[\k_1 ...\k_{11}]}
\int d^{16}\t~ 
\t^{\k_1} ... \t^{\k_{11}} f_{\a\b\g}(\t),}
which agrees with the tree amplitude prescription of \zpres\ up to a constant
normalization factor.

Note that the above computation can be easily generalized to correlation
functions where the picture-lowering operators are not at $y_I=\infty$.
In this case, one can get factors such as $\p \d(C\l)$ from OPE's
between the picture-lowering operators and the vertex operators.
However, since the amplitude is guaranteed to be independent of $C_I$,
one can use a similar argument to trivially perform the functional
integration over the pure spinor ghosts.

Using Lorentz invariance and symmetry properties,
the previous prescription for integrating over $\l^\a$ zero modes can
be generalized to
a prescription
for evaluating $\langle f(\l,C_I)\rangle$
where 
\eqn\genev{f(\l,C_I) = h(\l,C_I) \prod_{I=1}^{11} \p^{K_I} \d(C_I\l)}
and $h(\l,C_I)$ is a polynomial depending on $\l^\a$ and $C_{I\a}$ as
$$h(\l,C_I) = (\l)^{3+\sum_{I=1}^{11} K_I} \prod_{I=1}^{11}
(C_I)^{K_I+1}.$$
The manifestly Lorentz-covariant prescription is
\eqn\formgeral{\langle f(\l,C_I)\rangle =
\int[{\cal D} C]\int [{\cal D}\l] ~f(\l,C_I) }
$$=c'(\e{\cal T}^{-1})^{((\a\b\g))}_{[\rho_1 ...\rho_{11}]} 
{\p\over{\p\l^\a}}
{\p\over{\p\l^\b}}
{\p\over{\p\l^\g}}
{\p\over{\p C_{1\rho_1}}}
...
{\p\over{\p C_{11\rho_{11}}}} \prod_{I=1}^{11}
({\p\over{\p\l^\d}} {\p\over{\p C_{I\d}}})^{K_I} h(\l,C_I),$$
where $c'$ is a proportionality constant which can be computed as in \detc.

As will be shown in section 5, similar methods can be used to perform
functional integration over the $N^{mn}$ and $J$ zero modes in loop amplitudes.
However, before discussing these loop amplitudes, it will be necessary
to first construct an appropriate $b$ ghost.

\newsec{Construction of $b$ Ghost}

To compute $g$-loop amplitudes, the usual string theory prescription
requires the insertion of  $(3g-3)$ $b$ ghosts of $-1$ ghost-number 
which satisfy
\eqn\busual{\{Q,b(u)\}= T(u)}
where $T$ is the stress tensor of \stress. After integrating $b(u)$ with
a Beltrami differential $\mu_P(u)$ for $P=1$ to $3g-3$, the BRST
variation of $b(u)$ generates a total derivative with respect to the
Teichmuller parameter $\tau_P$ associated to the Beltrami differential $\mu_P$.
But since $w_\a$
can only appear in gauge-invariant combinations of zero ghost number,
there are no operators of negative ghost number in the pure spinor
formalism, so one cannot construct such a $b$ ghost.
Nevertheless, as will now be shown, the picture-raising operator 
$$Z_B = \half B_{mn}(\l\g^{mn}d) \d(BN)$$
can be used to construct a suitable
substitute
for the $b$ ghost in non-zero picture.

Since genus $g$ amplitudes also require $10g$ insertions of $Z_B(z)$,
one can combine $(3g-3)$ insertions of $Z_B(z)$ with the desired 
insertions of the $b(u)$ ghost and look for a non-local
operator $\widetilde b_B(u,z)$ which satisfies
\eqn\bhatnew{\{Q,\widetilde b_B(u,z)\} = T(u) Z_B(z).}
Note that $Z_B$ carries $+1$
ghost-number, so $\widetilde b_B$ carries zero ghost number.
And \bhatnew\ implies that integrating $\widetilde b(u,z)$ with the
Beltrami differential $\mu_P(u)$ has the same properties as 
integrating $b(u)$ with $\mu_P(u)$ in the presence of a picture-raising
operator $Z_B(z)$. 

Using 
$$Z_B(z) = Z_B(u) + \int_u^z dv\p Z_B(v) =
Z_B(u) + \int_u^z dv \{Q, B_{pq} \p N^{pq}(v) \d(BN(v))\},$$
one can define
\eqn\learns{\widetilde b_B(u,z) = b_B(u) + T(u)
\int_u^z dv B_{pq}\p N^{pq}(v) \d(BN(v))}
where $b_B(u)$ is a local operator satisfying\foot{
A similar picture-raised
version of the $b$ ghost 
appears in the N=4 topological description of the superstring
\ref\topo{N. Berkovits and C. Vafa, {\it N=4 Topological Strings},
Nucl. Phys. B433 (1995) 123, hep-th/9407190.} as
the $\widetilde G^-$ generator. Since the pure spinor formalism can
be related to the N=4 topological description through the twistor approach
of
\ref\tonin{M. Matone, L.
Mazzucato, I. Oda, D. Sorokin and M. Tonin, {\it The Superembedding
Origin of the Berkovits Pure Spinor Covariant Quantization of 
Superstrings}, Nucl. Phys. B639 (2002) 182, hep-th/0206104.}
\ref\twistors{
D. Sorokin, {\it Superbranes and Superembeddings}, Phys. Rep. 329 (2000) 1,
hep-th/9906142\semi
N. Berkovits, {\it The Heteterotic Green-Schwarz Superstring on
an N=(2,0) Worldsheet}, Nucl. Phys. B379 (1992) 96, hep-th/9201004\semi
M. Tonin, {\it World Sheet Supersymmetric Formulations of Green-Schwarz
Superstrings}, Phys. Lett. B266 (1991) 312\semi
D. Sorokin, V.I. Tkach, D.V. Volkov and A.A. Zheltukhin, {\it
{}From the Superparticle Siegel Symmetry to the Spinning Particle
Proper Time Supersymmetry}, Phys. Lett. B216 (1989) 302.}, it would
be interesting to try to relate $b_B$ with $\widetilde G^-$ using
the approach of \tonin.}
\eqn\bnew{\{Q,b_B(u)\} = T(u) Z_B(u).}

\subsec{Chain of operators}

To construct $b_B$ satisfying \bnew, it is useful to first
construct a chain of operators which are related to the stress tensor
$T$ through BRST transformations.
Although there is no $b$ operator of $-1$ ghost-number
satisfying $\{Q,b\}=T$,
there is an operator $G^\a$ of zero ghost-number satisfying
\eqn\Gnew{\{Q, G^\a\} = \l^\a T.} 
The existence of $G^\a$ is guaranteed since $\l^\a T$ is a BRST-invariant
operator of $+1$ ghost number and $+2$ conformal weight, and the
BRST cohomology at
$+1$ ghost number is non-trivial only at zero conformal weight.
One finds that\relating 
\eqn\GGdef{G^\a = \half \Pi^m (\g_m\ d)^\a - {1\over 4}
N_{mn} (\g^{mn}\p\t)^\a -{1\over 4} J\p\t^\a -{1\over 4}\p^2\t^\a}
where the $\p^2\t^\a$ term comes from normal-ordering.
Note that
if one ignores
this
normal-ordering contribution, all terms
in $G^\a$ carry ``engineering dimension'' $5\over 2$ where
$[\l^\a,\t^\a,x^m, p_\a, w_\a]$ are defined to carry engineering dimension
$[0,\half,1,{3\over 2},2 ]$.
Furthermore, one can verify that
$G^\a$ is a primary field of $+2$ conformal weight.

Since 
\eqn\gone{Q(\l^\a G^\b ) = \l^\a \l^\b T}
is symmetric and $\g$-matrix traceless 
(i.e. $Q(\l^{[\a} G^{\b]})= Q(\l\g^m G)=0$), cohomology arguments imply
there exists an operator $H^{\a\b}$ which satisfies
\eqn\gtwo{Q(H^{\a\b}) = \l^\a G^\b + g^{((\a\b))} }
where $g^{((\a\b))}$ is some symmetric $\g$-matrix traceless operator.\foot{
Since $(\l^\a G^\b -\l^{((\a} G^{\b))})$ is a BRST-invariant
operator of $+1$ ghost-number and $+2$ conformal weight, it is guaranteed
that 
$(\l^\a G^\b -\l^{((\a} G^{\b))})= Q(H^{\a\b}-H^{((\a\b))})$ for
some $H^{\a\b}$. Defining $g^{((\a\b))}=-\l^{((\a} G^{\b))} +
Q(H^{((\a\b))})$, one recovers \gtwo.}
Note that \gtwo\ only determines $H^{\a\b}$ up to the gauge transformation
\eqn\gthree{\d H^{\a\b} = \Omega^{((\a\b))} }
where $\Omega^{((\a\b))}$ is any symmetric $\g$-matrix traceless operator.
For example, one can choose $\Omega^{((\a\b))}$ such that $H^{((\a\b))}=0$,
in which case \gtwo\ is solved by 
\eqn\gfour{ H^{\a\b} ={1\over{16}} 
\g_m^{\a\b} ( N^{mn}\Pi_n -\half J\Pi^m + 2\p\Pi^m)
 + {1\over{384}}\g_{mnp}^{\a\b} (d\g^{mnp} d 
+24 N^{mn}\Pi^p)}
where $g^{((\a\b))}=-{1\over{3840}} \g_{mnpqr}^{\a\b} (\l\g^{mnpqr} G).$
One can check that $H^{\a\b}$ is a primary field of conformal weight $+2$
and that, if one ignores
the normal-ordering term proportional to $\p\Pi^m$,
all terms in $H^{\a\b}$ carry $+3$ engineering dimension.\foot{It is
interesting to note that the $[T,G^\a, H^{\a\b}]$ operators closely
resemble the $[A,B^\a, C^{mnp}]$ constraints of Siegel \siegel\ 
for quantization
of the superparticle and superstring.}

The next link in the chain of operators is constructed by noting that
\eqn\hone{Q (\l^\a H^{\b\g}) = \l^\a \l^\b G^\g + \l^\a g^{((\b\g))},}
which implies using similar cohomology arguments as before
that there exists an operator $K^{\a\b\g}$ which satisfies
\eqn\htwo{ 
Q( K^{\a\b\g} ) = \l^\a H^{\b\g} + h_1^{((\a\b))\g} + h_2^{\a((\b\g))} }
where $h_1^{((\a\b))\g}$ is some operator which is symmetric $\g$-matrix
traceless in its first two indices, and 
$h_2^{\a((\b\g))}$ is some operator which is symmetric $\g$-matrix
traceless in its last two indices. Note that \htwo\ only
determines $K^{\a\b\g}$ up to the gauge transformation
\eqn\hthree{\d K^{\a\b\g} = \Omega_1^{((\a\b))\g} + \Omega_2^{\a((\b\g))}. }
{}From its ${7\over 2}$ engineering dimension (ignoring normal-ordering
terms) and its $+2$ conformal weight, one can deduce that 
\eqn\hfour{K^{\a\b\g} = c_{1 mn}^{\a\b\g\rho} N^{mn} d_\rho +
 c_2^{\a\b\g\rho} J d_\rho +
 c_3^{\a\b\g\rho} \p d_\rho,}
however, the coefficients $[
c_{1 mn}^{\a\b\g\rho},
c_2^{\a\b\g\rho}, c_3^{\a\b\g\rho}]$ have not yet been computed.\foot
{For this type of computation, it would be very helpful to have a
computer code designed to handle manipulations of pure spinors.}

Finally, the last link in the chain of operators is constructed by
noting that
\eqn\kone{
Q (\l^\a K^{\b\g\d}) = \l^\a \l^\b H^{\g\d} + \l^\a h_1^{((\b\g))\d}
+\l^\a h_2^{\b((\g\d))},}
which implies that there exists an operator $L^{\a\b\g\d}$ which
satisfies
\eqn\ktwo{ 
Q ( L^{\a\b\g\d} ) = \l^\a K^{\b\g\d} + k_1^{((\a\b))\g\d} + 
k_2^{\a((\b\g))\d}  +
k_3^{\a\b((\g\d))} ,}
where $[k_1^{((\a\b))\g\d},k_2^{\a((\b\g))\d},k_3^{\a\b((\g\d))}]$
are operators which are symmetric $\g$-matrix traceless in their
first two, middle two, or last two indices. As before, \ktwo\ only
determines $L^{\a\b\g\d}$ up to the gauge transformation
\eqn\kthree{
\d L^{\a\b\g\d} = \Omega_1^{((\a\b))\g\d} + \Omega_2^{\a((\b\g))\d}
+ \Omega_3^{\a\b((\g\d))}.}
Since $L^{\a\b\g\d}$ carries $+4$ engineering dimension (ignoring
normal-ordering terms) and $+2$ conformal weight, it has the form
\eqn\kfour{
L^{\a\b\g\d} = c_{4 mnpq}^{\a\b\g\d} N^{mn} N^{pq} +
 c_{5 mn}^{\a\b\g\d} J N^{mn} +
 c_6^{\a\b\g\d} JJ + c_{7 mn}^{\a\b\g\d} \p N^{mn} +  
 c_{8 }^{\a\b\g\d} \p J, }
where the coefficients in \kfour\ have not yet been computed.

To show that $L^{\a\b\g\d}$ is the last link in the chain of operators,
note that there are no supersymmetric primary fields of $+2$
conformal weight which carry engineering dimension greater than four. 
So if one tries to define an operator $M^{\a\b\g\d}$ satisfying
\eqn\lone{Q (M^{\a\b\g\d\rho}) = \l^\a L^{\b\g\d\rho} + l_1^{((\a\b))\g\d\rho}
+ l_2^{\a((\b\g))\d\rho}
+ l_3^{\a\b((\g\d))\rho}
+ l_4^{\a\b\g((\d\rho))}}
for some $[l_1^{((\a\b))\g\d\rho},
l_2^{\a((\b\g))\d\rho},l_3^{\a\b((\g\d))\rho},l_4^{\a\b\g((\d\rho))}]$,
one finds that $M^{\a\b\g\d\rho}$ must vanish. This implies that
\eqn\ltwo{
\l^\a L^{\b\g\d\rho} = - l_1^{((\a\b))\g\d\rho}
- l_2^{\a((\b\g))\d\rho}
- l_3^{\a\b((\g\d))\rho}
- l_4^{\a\b\g((\d\rho))}, }
which implies that 
\eqn\lthree{ L^{\a\b\g\d} = \l^\a S^{\b\g\d} + s_1^{((\a\b))\g\d}
+ s_2^{\a((\b\g))\d}
+ s_3^{\a\b((\g\d))}}
for some $S^{\b\g\d}$ and 
$[ s_1^{((\a\b))\g\d},
 s_2^{\a((\b\g))\d},
s_3^{\a\b((\g\d))}]$. 
For the following subsection, it will be useful to note that 
\lthree\ and \ktwo\ imply that 
\eqn\identityS{Q ( S^{\b\g\d} ) = K^{\b\g\d} +
\l^\b T^{\g\d} + t_1^{((\b\g))\d} + t_2^{\b((\g\d))}}
for some $[T^{\g\d}, t_1^{((\b\g))\d}, t_2^{\b((\g\d))}].$

Note that $S^{\b\g\d}$ has ghost-number $-1$, so
it will depend on $w_\a$ in combinations which are not invariant under
the gauge transformation of \gaugew. However, since $L^{\a\b\g\d}$ only
involves gauge-invariant combinations of $w_\a$, the change in 
$S^{\b\g\d}$ under \gaugew\ must be of the form
\eqn\schange{\d S^{\b\g\d} = \l^\b \Sigma^{\g\d} +
\rho_1^{((\b\g))\d} 
+\rho_2^{\b((\g\d))} }
for some $[\Sigma^{\g\d}, \rho_1^{((\b\g))\d},
\rho_2^{\b((\g\d))}]$ in order that the change in $S^{\b\g\d}$ can
be cancelled in \lthree\ by shifting 
\eqn\shifting{\d s_1^{((\a\b))\g\d} = -\l^\a\l^\b \Sigma^{\g\d},\quad
\d s_2^{\a((\b\g))\d} = -\l^\a \rho_1^{((\b\g))\d},\quad
\d s_3^{\a\b((\g\d))} = -\l^\a \rho_2^{\b((\g\d))}.}

\subsec{ Construction of $b_B$}

Since $[Q, TZ_B]=0$ and $T Z_B$ has $+1$ ghost-number and $+2$
conformal weight, cohomology arguments\foot
{At zero picture and $+1$ ghost-number,  
the BRST cohomology is trivial for states of nonzero conformal weight.
It is expected that this is also true at nonzero picture, however, this has
not yet been verified.}
suggest one can find an operator
$b_B$ satisfying $\{Q,b_B\}=TZ_B$.
Although the structure of $b_B$ will be complicated, one can construct
$b_B$ iteratively using the operators $[T,G^\a,H^{\a\b},K^{\a\b\g},
L^{\a\b\g\d}]$ of the previous
subsection. To construct $b_B$, first note that 
\eqn\Teq{T Z_B = \half T (\l Bd) \d(BN) =
\half \{Q,G^\a\} (Bd)_\a  \d(BN) }
$$=\half
\{Q,(G\g^{mn}d) B_{mn} \d(BN) \} + \half
G^\a [Q,(Bd)_\a \d(BN)] $$
$$= 
\{Q, b_B^{(1)}\} +\half G^\a[-(\g^{mn}\g^p\l)_\a B_{mn} \Pi_p \d(BN) -
\half(Bd)_\a (\l Bd)  \p\d(BN)]$$
where 
\eqn\defbone{b_B^{(1)} = \half
(G\g^{mn}d) B_{mn} \d(BN),}
$(Bd)_\a \equiv (\g^{mn} d)_\a B_{mn}$, $(\l Bd) \equiv (\l\g^{mn} d) B_{mn}$,
and normal-ordering contributions are being ignored.

So one now needs to find an operator $b_B - b^{(1)}_B$ which satisfies
\eqn\Geq {\{Q, b_B-b^{(1)}_B\} = - 
G^\a[\half (\g^{mn}\g^p\l)_\a B_{mn} \Pi_p \d(BN) +{1\over 4}
(Bd)_\a 
(\l Bd) \p\d(BN)]}
$$= -\{Q, H^{\b\a}[\half (\g^p\g^{nm})_{\b\a}\Pi_p B_{mn} \d(BN)
+{1\over 4} (Bd)_\a (Bd)_\b \p \d(BN)]\}$$
$$+ H^{\b\a} \{ Q,
\half (\g^p\g^{nm})_{\b\a}\Pi_p B_{mn} \d(BN)
+{1\over 4} (Bd)_\a (Bd)_\b \p \d(BN)\}$$
$$= \{Q, b_B^{(2)}\} + H^{\b\a}
[
\half (\g^p\g^{nm})_{\b\a}(\l\g_p\p\t) B_{mn} \d(BN)
+{1\over 4} (\g^p\g^{nm})_{\b\a}\Pi_p B_{mn}(\l Bd)\p\d(BN)$$
$$
+{1\over 4} (Bd)_{[\a} (\g_{mn}\g_p)_{\b]} B^{mn} \Pi^p \p \d(BN)
+{1\over 8} (Bd)_\a (Bd)_\b (\l B d)\p^2 \d(BN)]$$
where 
\eqn\defbtwo{b_B^{(2)} = 
-H^{\b\a} 
[\half (\g^p\g^{nm})_{\b\a}\Pi_p B_{mn} \d(BN)
+{1\over 4} (Bd)_\a (Bd)_\b \p \d(BN)].}

One now continues this procedure two more stages to construct
$b_B^{(3)}$ using $K^{\a \b\g}$, and to construct
$b_B^{(4)}$ using $L^{\a\b\g\d}$ and $S^{\b\g\d}$. 
Using the properties of \lthree\
and \identityS,
one can verify that the  
procedure stops here and, ignoring normal-ordering
contributions,
\eqn\stopshere{b_B = b_B^{(1)} + b_B^{(2)} + b_B^{(3)} + b_B^{(4)}}
where
$b_B^{(1)}$ and $b_B^{(2)}$ are given in \defbone\ and \defbtwo, and
$$b_B^{(3)} = \half K^{\g\b\a}[(\g^p\g^{mn})_{\b\a}(\g_p\p\t)_\g
B_{mn}\d(BN) +$$
$$+ {1\over 2}(\g^p\g^{nm})_{\b\a} (Bd)_\g \Pi_p B_{mn} \p\d(BN)
+ {1\over 2}(\g^p\g^{nm})_{\g[\b} (Bd)_{\a]} \Pi_p B_{mn}  \p\d(BN)$$
$$
+{1\over 4}(Bd)_\a (Bd)_\b (Bd)_\g \p^2\d(BN)]
, $$
$$b_B^{(4)} = 
\half S^{\g\b\a} (\g^p\g^{nm})_{\b\a} (\g_p\p\l)_\g B_{mn} \d(BN)$$
$$ +{1\over 4}
L^{\d\g\b\a}[ ~
( ~
(\g^p\g^{nm})_{\b\a} (Bd)_{[\d} (\g_p\p\t)_{\g]} 
-(\g^p\g^{nm})_{\g[\b} (Bd)_{\a]} (\g_p\p\t)_{\d}~)~ B_{mn} \p\d(BN) $$
$$ - (~(\g^s\g^{rq})_{\d[\g}
(\g^p\g^{nm})_{\b]\a} 
+ (\g^s\g^{rq})_{\d\a} (\g^p\g^{nm})_{\g\b} ~) ~
\Pi_p B_{mn} \Pi_s B_{qr} \p\d(BN)
$$
$$ -{1\over 2}
(~ 
(\g^p\g^{nm})_{\b\a} (Bd)_\g (Bd)_\d +
(\g^p\g^{nm})_{\g[\b} (Bd)_{\a]} (Bd)_\d  $$
$$ +
\half
(\g^p\g^{nm})_{\d[\a} (Bd)_{\b} (Bd)_{\g]} ~)~
\Pi_p B_{mn} \p^2\d(BN)$$
$$-{1\over{4}}(Bd)_\a (Bd)_\b (Bd)_\g (Bd)_\d \p^3\d (BN)~].$$

Although $b_B$ of \stopshere\ is a complicated operator, it has certain simple
properties which will be useful to point out. Firstly, 
$b_B$ is invariant under the gauge transformations
of \gthree, \hthree, \kthree\ and \schange\ 
for $H^{\a\b}$, $K^{\a\b\g}$, $L^{\a\b\g\d}$ and $S^{\b\g\d}$.
Secondly, all terms in
$b_B$ have $+2$ conformal weight (where $\d(BN)$ has $-1$ conformal
weight). Thirdly, if one
ignores normal-ordering contributions\foot{Since terms coming
from normal-ordering carry engineering dimension less than $+4$,
they will not contribute to the scattering amplitudes computed in
section 6. However, for more general amplitude computations, one will
need to include contributions from the normal-ordering terms in $b_B$.},
all terms
in $b_B$ have $+4$ engineering dimension where 
$[\l^\a,\t^\a,x^m, d_\a, w_\a]$ carry engineering dimension
$[0,\half,1,{3\over 2}, 2]$ and $\d(BN)$ is defined
to carry zero engineering dimension\foot{Although
it might seem more natural to
define $\d(BN)$ to carry $-2$ engineering dimension, it will be more
convenient for our purposes to define $\d(BN)$ to be dimensionless.}. 
Fourthly, all terms in $b_B$ are manifestly spacetime supersymmetric.
And finally, although $b_B$ is not Lorentz-invariant, its Lorentz
transformation only affects the scattering amplitude by a surface term.

To verify this last statement, 
note that under Lorentz transformations generated by
$M^{mn}$, \mzb\ implies that $M^{mn} Z_B = Q\Lambda_B^{mn}$ where
\eqn\deflambda{\Lambda_B^{mn} =2 \eta^{p[m}\eta^{n]}_r B_{pq} N^{qr} \d(BN).}
Since $\{Q,b_B\}= TZ_B$, this implies that 
$$M^{mn} b_B = T \Lambda_B^{mn} + Q\Omega^{mn}_B$$
for some $\Omega^{mn}_B$. 
So using \learns, 
\eqn\transbhat{M^{mn}\widetilde b_B(u,z) = 
M^{mn} b_B(u) +  T(u) \int_u^z dv M^{mn}(B_{pq} \p N^{pq} \d(BN))}
$$
=T(u)\Lambda^{mn}_B(u) + Q\Omega^{mn}_B(u)
+T(u)\int_u^z dv \p\Lambda_B^{mn}(v) $$
$$ = T(u) \Lambda_B^{mn}(z) + Q\Omega^{mn}_B(u).$$
Since 
$T(u) \Lambda_B^{mn}(z)$
produces a total derivative with respect to the Teichmuller
parameter $\tau_P$ associated
to the Beltrami differential $\mu_P(u)$, the Lorentz variation of $b_B$
only changes the scattering amplitude by a surface term. 

\newsec{Multiloop Amplitude Prescription}

Using the picture-changing operators of section 3 and the $b_B$ ghost
of section 4, one can give a super-Poincar\'e covariant prescription
for computing $N$-point $g$-loop closed superstring scattering amplitudes as
\eqn\gloop{{\cal A} = 
\int d^2\tau_1 ...d^2\tau_{3g-3} \langle ~ |~
\prod_{P=1}^{3g-3}\int d^2 u_P \mu_P(u_P)
\widetilde b_{B_P}(u_P,z_P) }
$$
\prod_{P=3g-2}^{10g} Z_{B_P}(z_P) \prod_{R=1}^{g} Z_J(v_R)
\prod_{I=1}^{11} Y_{C_I}(y_I)~|^2 ~\prod_{T=1}^N \int d^2 t_T U_T(t_T)
~\rangle,$$
where $|~~|^2$ signifies the left-right product, 
$\tau_P$ are the Teichmuller parameters associated to the 
Beltrami differentials $\mu_P (u_P)$, and $U_T(t_T)$ are the
dimension $(1,1)$ closed string
vertex operators for the $N$ external states. 
The constant antisymmetric tensors $B_P^{mn}$ in $b_{B_P}$ and
$Z_{B_P}$ will be chosen to satisfy \Bconv\ and will also be chosen such that
$B_I=B_{I+10} = ... = B_{I+10(g-1)}$ for $I=1$ to 10. In other words, there
will be ten constant antisymmetric
tensors $B_I^{mn}$, each of which appear in $g$ picture-raising
operators or $b_B$ ghosts. One possible choice for these ten tensors
is $B_{I~ mn} N^{mn} =\d_{I~[ab]} N^{[ab]}$ where $a,b=1$ to 5, $[ab]$
is a ten-component representation of $SU(5)$, and
$N^{[ab]}$ transforms as a $10$
representation under the SU(5) subgroup of the (Wick-rotated)
Lorentz group SO(10).\foot{In terms of the
U(5)-covariant variables of \superp, this choice would imply 
$B_{I~mn} N^{mn}=\d_{I~[ab]}v^{[ab]}$.} 

When $g=1$, the prescription of \gloop\ needs to be modified for the 
usual reason that
genus-one worldsheets are invariant under constant translations, so
one of the vertex operators should be unintegrated.
The one-loop amplitude prescription is therefore
\eqn\oneloop{{\cal A} = 
\int d^2\tau \langle ~|~
\int d^2 u \mu(u)
\widetilde b_{B_1}(u,z_1) }
$$
\prod_{P=2}^{10} Z_{B_P}(z_P) Z_J(v)
\prod_{I=1}^{11} Y_{C_I}(y_I)~|^2 ~V_1(t_1)
\prod_{T=2}^N \int d^2 t_T U_T(t_T)~\rangle,$$
where $V_1(t_1)$ is the unintegrated 
closed string
vertex operator.

As shown in the previous sections, the Lorentz variations of $\widetilde
b_{B_P}$,
$Z_{B_P}$ and $Y_{C_I}$ are BRST-trivial, so the prescriptions of \gloop\ 
and \oneloop\
are Lorentz-invariant up to possible surface terms. Also, all operators
in \gloop\ and \oneloop\
are manifestly spacetime supersymmetric except for $Y_{C_I}$,
whose supersymmetry variation is BRST-trivial. In section 6, it will
be argued that surface terms can be ignored in this formalism because
of finiteness properties of the correlation functions. So the 
amplitude prescriptions of \gloop\ and \oneloop\
are super-Poincar\'e covariant. This
implies that ${\cal A}$ is independent of the eleven constant spinors $C_I$ and
ten constant tensors $B_P$ which appear in the picture-changing operators.
As will now be shown, functional integration over the matter fields and
pure spinor ghosts can be used to derive manifestly Lorentz-covariant
expressions from the amplitude prescriptions of \gloop\ and \oneloop.

As usual, the functional integration factorizes into partition
functions and correlation functions for the different worldsheet variables.
However, in the pure spinor formalism, the partition functions for the
different worldsheet
variables cancel each other out. This is easy to verify since
the partition function for the ten bosonic $x^\mu$ variables gives a factor of
$(\det \bar\p_0)^{-5}
(\det \p_0)^{-5}$ where $\p_0$ and $\bar\p_0$ are the holomorphic
and antiholomorphic derivatives acting on fields of zero conformal weight,
the 
partition function for the sixteen fermionic $(\t^\a,p_\a)$ and
$(\bar \t^\a,\bar p_\a)$ variables gives a factor of 
$(\det \bar\p_0)^{16}
(\det \p_0)^{16}$, and the partition
function for the eleven bosonic $(\l^\a,w_\a)$ and
$(\bar \l^\a,\bar w_\a)$ variables gives a factor of 
$(\det \bar\p_0)^{-11}
(\det \p_0)^{-11}$.
So to perform the functional integral, one only needs to compute the
correlation functions for the matter variables and pure spinor 
ghosts. 

\subsec{Correlation function for matter variables}

In computing the $g$-loop correlation functions, 
one can follow the same general
procedure as in the tree amplitude computation of
subsection (3.3),
but one now needs to take into account the $g$ zero modes of the fields
with $+1$ conformal weight. For example,
one can functionally integrate over the $(\t^\a,p_\a)$ variables
by first separating off the zero mode of $d_\a$ by writing
\eqn\dzerom{d_\a(z) = \sum_{R=1}^g d_\a^R \om_R(z) +\widehat d_\a(z)}
where $\om_R$ are the $g$ holomorphic one-forms and $d_\a^R$
are the $16g$ zero modes of $d_\a$. Since the poles
of $\widehat d_\a(z)$ are determined by its OPE's with the other
fields, these OPE's completely fix the dependence of the correlation
function on the location $z$. 

For example, the $g$-loop analog of the correlation function of \corex\ is
given by
\eqn\corex{\langle
d_\a(z) \Pi^m(u) \p\t^\b(v) d_\g(w) A_\d(x(y),\t(y))\rangle}
$$
=d^R_\a \om_R(z) \langle\Pi^m(u) \p\t^\b(v) d_\g(w) A_\d(x(y),\t(y))\rangle$$
$$+
\g^m_{\a\rho} F(z,u) \langle\p\t^\rho(u)
\p\t^\b(v) d_\g(w) A_\d(x(y),\t(y))\rangle$$
$$+\d_\a^\b
\p_v F(z,v) \langle\Pi^m(u) d_\g(w) A_\d(x(y),\t(y))\rangle$$
$$+\g_{\a\g}^n F(z,w)
\langle\Pi^m(u)\p\t^\b(v) \Pi^n(w) A_\d(x(y),\t(y))\rangle$$
$$+ F(z,y)
\langle\Pi^m(u)\p\t^\b(v) \d_\g(w) D_\a A_\d(x(y),\t(y))\rangle$$
where
$$F(z,y) = \p_z \log E(z,y)$$
and $E(z,y)$ is the holomorphic prime form which goes like
$(z-y)$ when $z$ approaches $y$ \ref\dreview{E. D'Hoker and D. Phong,
{\it The Geometry of String Perturbation Theory}, Reviews of Modern
Physics, Vol. 60 (1988) 917.}\ref\vv{E. Verlinde and H. Verlinde,
{\it Chiral Bosonization, Determinants and the String Partition
Function}, Nucl. Phys. B288 (1987) 357.}.

Using this procedure, one can remove the $d_\a$'s one at a time
from the correlation function. After all the $d_\a$'s have been
removed, one can replace all remaining $\t$'s in the correlation function
by their zero mode. One then functionally integrates over the 16 
$\t^\a$ zero modes and the $16g$ $d^R_\a$ zero modes using standard
Berezin integration. 

Although the functions $F(z,y)$ coming from the
$\widehat d_\a$ OPE's in \corex\
are not single-valued when either $z$ or $y$
goes around a $B$-cycle of the genus $g$ surface, the scattering
amplitude will be single-valued.
This is because when $d_\a(z)$ goes around the $R^{th}$ $B$-cycle,
$F(z,y)\to F(z,y) -2\pi i \om_R(z)$,
which induces a change in the correlation function of \corex\ by the term
\eqn\changecorex{-2\pi i \om_R(z)
\langle\oint ds d_\a(s)~ \Pi^m(u) \p\t^\b(v) d_\g(w) A_\d(x(y),\t(y))\rangle}
where the contour integral of $s$ goes around all the other points on
the surface. Since $d_\a$ is a conserved current, the contour integral
can be deformed off the back of the surface, giving no contribution
to the scattering amplitude. 

And when $\t^\a(y)$ goes around the 
$R^{th}$ $B$-cycle, 
$F(z,y)\to F(z,y) +2\pi i \om_R(z)$. Since this change in $F(z,y)$
is independent
of $y$ and is proportional to $\om_R(z)$, 
the resulting change in the $d_\a$ correlation function
can be cancelled by shifting the $d_\a^R$ zero mode in \dzerom\
by a $z$-independent amount. Since
Berezin integration is unchanged by a constant shift of the Grassmann 
variables, the scattering amplitude is single-valued after integration
over the $d_\a^R$ zero mode.

After performing functional integration over the $(\t^\a,p_\a)$ variables
in this manner,
one can easily perform functional integration over the $x^m$
variables using the standard techniques \dreview\vv. 
For example, the correlation
function
$\langle\prod_{r=1}^N \exp(i k\cdot x(u_r)\rangle$ is equal to
\eqn\xcorr{
=\prod_{R=1}^g \int d^{10} P_R ~|~\exp (i\pi P_R\cdot P_S \tau_{RS} 
+2\pi i \sum_{r=1}^N k_r\cdot P_R\int^{u_r} dv \om_R (v))
\prod_{r<s} E(u_r,u_s)^{k_r\cdot k_s}~|^2}
where $P^m_R$ is the loop momentum through the $R^{th}$ $A$-cycle,
$\tau_{RS}$ is the period matrix, and $E(u,v)$ is the holomorphic
prime form. 

\subsec{Correlation function for the pure spinor ghosts}

After functionally integrating over the matter variables, one is left with
a correlation function depending on the pure spinor ghost operators
$\l^\a$, $N^{mn}$ and $J$. To compute this correlation function, first
separate off the $g$ zero modes of $N_{mn}$ by writing
\eqn\Nmnz{N_{mn}(z) = N_{mn}^R \om_R(z) + \widehat N_{mn}(z).}
Since the singularities of $\widehat N_{mn}(z)$ are determined from the
OPE's of \OPE, the dependence of the correlation function on $z$ is
completely determined. 

For example, a $g$-loop analog of the computation of \coren\ is
\eqn\ncorre{
\langle N_{mn}(z) N_{pq}(u) \l^\a(v) \d(BN(w)) \d(C\l(y))\rangle =}
$$= N_{mn}^R \om_R(z) \langle 
N_{pq}(u) \l^\a(v) \d(BN(w)) \d(C\l(y))\rangle $$
$$+ F(z,u)
\langle (\eta_{p[n} N_{m]q}(u) -\eta_{q[n} N_{m]p}(u)) \l^\a(v)
\d(BN(w)) \d(C\l(y))\rangle $$
$$-3\p_u F(z,u) \langle \eta_{q[m}\eta_{n]p} \l^\a(v)
\d(BN(w)) \d(C\l(y))\rangle $$
$$+\half F(z,v) \langle N_{pq}(u) (\g_{mn}\l(v))^\a
\d(BN(w)) \d(C\l(y))\rangle $$
$$+ 2 F(z,w) \langle N_{pq}(u) \l^\a(v)
B_{r[m} N_{n]}{}^r(w)\p\d(BN(w)) \d(C\l(y))\rangle $$
$$+6 \p_w F(z,w) \langle N_{pq}(u) \l^\a(v)
B_{mn}\p\d(BN(w)) \d(C\l(y))\rangle $$
$$+\half F(z,y) \langle N_{pq}(u) \l^\a(v)
\d(BN(w)) (C\g_{mn}\l(y))\p \d(C\l(y))\rangle .$$

If one counts $\p^L \d(BN)$ as containing $(-L)$ $N$'s, then the number of $N$'s
is decreased after performing this correlation function. So repeating
this procedure enough times will eventually give a correlation function with
a net zero number of $N$'s, at which point one can stop. Note that the
procedure of separating off the zero mode of $N_{mn}(z)$ must also
be used for the $N_{mn}$ appearing in $\d(BN)$. So one needs to include
the contribution from 
\eqn\contrdn{\d(B N(z)) =\d(B N^R \om_R(z) + B \widehat N(z))}
$$= \d(B N^R \om_R(z)) + (B\widehat N(z))\p\d(B N^R\om_R(z))
+ \half (B\widehat N(z))^2 \p^2 \d(B N^R\om_R(z)) + ... ,$$
where one uses the OPE of $\widehat N(z)$ 
with the other fields
to determine the dependence of the correlation
function on $z$. 

As in the $(\t^\a,p_\a)$ correlation function, although 
$F(z,y)$ is not single-valued when either $z$ or $y$ goes around
a $B$-cycle, the scattering amplitude will be single-valued. When
$N_{mn}(z)$ goes around the $R^{th}$ $B$-cycle, the change in the
correlation function of \ncorre\ is equal to 
\eqn\changenc{-2\pi i\om_R(z)\langle (\oint ds 
N_{mn}(s)) N_{pq}(u) \l^\a(v) \d(BN(w)) \d(C\l(y))\rangle }
where the contour integral of $s$ goes around all points on the surface.
Since $N_{mn}$ is a conserved current, the contour can be deformed off the
surface, so this contribution vanishes. And when $\l^\a(y)$ goes around a
$B$-cycle, the change in the correlation function is independent of $y$
and can be cancelled by an appropriate shift of the $N_{mn}^R$ zero modes.
So after integrating over the $N_{mn}^R$ zero modes using a shift-invariant 
measure, this contribution will also vanish. 

After removing all the $N_{mn}$'s from the correlation function and replacing
them with $N_{mn}^R$ zero modes, one can follow the same procedure for the
$J(z)$'s in the correlation function. For example, after separating off
the $g$ zero modes by writing 
$$J(z) = J^R\om_R(z) + \widehat J(z),$$
one
can use the OPE's of \OPE\ to show that 
\eqn\Jcorr{\langle J(z)  \d(J(u))\l^{\a_1}(v_1) ...\l^{\a_M}(v_M) 
\d(C_1\l(y_1)) ...\d(C_{11}\l(y_{11}))\rangle}
$$= 
J^R\om_R(z) \langle \d(J(u))\l^{\a_1}(v_1) ...\l^{\a_M}(v_M) 
 \d(C_1\l(y_1)) ...\d(C_{11}\l(y_{11}))\rangle$$
$$-4\p_u F(z,u)  
\langle \p\d(J(u))\l^{\a_1}(v_1) ...\l^{\a_M}(v_M) 
 \d(C_1\l(y_1)) ...\d(C_{11}\l(y_{11}))\rangle$$
$$+ (\sum_{Q=1}^M F(z,v_Q) - \sum_{I=1}^{11} F(z,y_I) + 8 \p_z(\ln\s(z)))$$
$$
\langle  \d(J(u))\l^{\a_1}(v_1) ...\l^{\a_M}(v_M) 
\d(C_1\l(y_1)) ...\d(C_{11}\l(y_{11}))\rangle$$
where the term 
proportional to $8 \p_z(\ln\s(z))$ comes from OPE's
with the screening charge which
is responsible for the ghost-number anomaly. As discussed in \vv, 
$\s(z)$ is a multivalued holomorphic function without zeros or poles
which satisfies
\eqn\changes{\p_z\ln\s(z) \to \p_z\ln\s(z) + 2\pi i (g-1) \om_R(z)}
when $z$ goes around the $R^{th}$ $B$-cycle. 
A convenient representation for 
$\p_z(\ln\s(z))$ is \verltwo
\eqn\convchoice{\p_z(\ln\s(z))= \sum_{R=1}^g \oint_{A_R} dv \om_R(v)
F(z,v)}
where $F(z,v) = \p_z \ln E(z,v)$ and $E(z,v)$ is the holomorphic prime form.

One can easily use \changes\ and the ghost-number anomaly to show that 
\Jcorr\ is invariant when $J(z)$ goes around the $R^{th}$ $B$ cycle.\foot
{Because of the ghost-number anomaly, $J$ is not a conserved
current and deforming $\oint ds J(s)$
off the surface gives a contribution which is cancelled by
\changes.} And when
$\l^\a(v)$ goes around the $R^{th}$ $B$-cycle, the change in the correlation
function can be cancelled by shifting the zero mode of $J^R$. 

After removing all $N$'s and $J$'s from the correlation function and replacing
them with $N_{mn}^R$ and $J^R$ zero modes, one can also replace all
remaining $\l^\a$'s in the correlation function by their zero mode. As
will now be described, one then
needs to integrate over the $(\l^\a,N_{mn}^R, J^R)$ zero modes using
the measure factors defined in subsection (3.1).

\subsec{Integral over pure spinor zero modes}

After integrating out the non-zero modes of $(\l^\a, N_{mn},J)$, one
obtains an expression 
$\langle f(\l,N_R,J_R,C_I,B_P) \rangle$ depending only on the zero modes
of $(\l^\a,N_{mn}^R,J^R)$ and the 
constant spinors and tensors $C_I$ and $B_P^{mn}$. The scattering
amplitude is then defined by the integral
\eqn\finteg{{\cal A}=\int [{\cal D}\l] [{\cal D} N_1] ... [{\cal D} N_g]
f(\l^\a, N_{mn}^R, J^R, C^I, B_{mn}^P)}
where $[{\cal D}\l]$ and $[{\cal D}N]$ are defined in \mld\ and \mlN.

Using the properties of the measure factors $[{\cal D}\l]$ and
$[{\cal D} N]$, one can write 
\eqn\fcan{{\cal A} = \int [{\cal D}\l] \prod_{R=1}^g [{\cal D} N_R]
f = \int (d^{11}\l)^{[\a_1 ... \a_{11}]}\prod_{R=1}^g
(d^{11}N_R)^{[[m_1^R n_1^R]...[m_{10}^R n_{10}^R]]}}
$$
\prod_{R=1}^g
(\g_{m_1^R n_1^R m_2^R m_3^R m_4^R})^{\rho_1^R \rho_2^R}
(\g_{m_5^R n_5^R n_2^R m_6^R m_7^R})^{\rho_3^R \rho_4^R}
(\g_{m_8^R n_8^R n_3^R n_6^R m_9^R})^{\rho_5^R \rho_6^R}
(\g_{m_{10}^R n_{10}^R n_4^R n_7^R n_9^R})^{\rho_7^R \rho_8^R}
$$
$$
(\e{\cal T}^{-1})^{((\b_1\b_2\b_3))}_{[\a_1 ...\a_{11}]} 
f_{((\b_1\b_2\b_3\rho_1^1 ...\rho_8^1 ...\rho_1^g ...\rho_8^g))}
(\l,N_R,J_R,C_I,B_P)$$
where 
$$f = \l^{\b_1}\l^{\b_2}\l^{\b_3} \l^{\rho_1^1} ... \l^{\rho_8^1}
... \l^{\rho_1^g} ... \l^{\rho_8^g}
f_{((\b_1\b_2\b_3\rho_1^1 ...\rho_8^1 ...\rho_1^g ...\rho_8^g))}
(\l,N_R,J_R,C_I,B_P).$$

As in the discussion of subsection (3.3) for tree amplitudes, \fcan\
is in general
a complicated function of the $B_P$'s and $C_I$'s.
However, using the properties of the picture-changing operators and
$b_B$ ghost, one knows that the scattering amplitude must be independent
of these constant spinors and tensors. One can therefore integrate
\fcan\ over all choices of $B_P$ and $C_I$ using a measure factor
$[{\cal D}B][{\cal D}C]$ which satisfies
$\int[{\cal D}B][{\cal D}C]=1$. Note that \fcan\ is manifestly
invariant under rescalings of $C_{I\a}$  
and $B_P^{mn}$, so these constant spinors and tensors can be
interpreted as projective variables.

Using arguments similar to those
of subsection (3.3), a manifestly Lorentz-covariant prescription will now
be given for evaluating $\langle f(\l,N_R,J_R,C_I,B_P)\rangle$. 
To be non-vanishing and have ghost-number $8g-8$, 
$f(\l,N_R,J_R,C_I,B_P)$ will depend on $(\l,N_R,J_R,C_I,B_P)$ as
\eqn\fdepend{f(\l,N_R,J_R,C_I,B_P) =}
$$ h(\l,N_R,J_R,C_I,B_P)\prod_{R=1}^g \p^{M_R} \d(J_R)
 \prod_{P=1}^{10} \prod_{R=1}^g \p^{L_{P,R}}
\d(B_P N_R))   \prod_{I=1}^{11} \p^{K_I} (C_I \l) ,$$
where $h$ is a polynomial depending on 
$(\l^\a,N_{mn}^R,J^R, C_{I\a}, B_P^{mn})$ as
\eqn\fprime{ (\l)^{8g-8 + \sum_{I=1}^{11} (K_I +1)}
\prod_{R=1}^g (J_R)^{M_R}
(N_R)^{\sum_{P=1}^{10} L_{P,R}}  
\prod_{P=1}^{10} (B_P)^{\sum_{R=1}^g (L_{P,R} +1)}
\prod_{I=1}^{11} (C_I)^{K_I+1}.}
Using Lorentz invariance and symmetry properties, one can argue
that 
\eqn\finalf{{\cal A} = \int [{\cal D}B][{\cal D}C]
\int [{\cal D}\l] \prod_{R=1}^g [{\cal D}N_R] f(\l,N_R,J_R,C_I,B_P)}
$$ =c' ~[~
({\p\over{\p\l}}
\g^{m_1 n_1 m_2 m_3 m_4}
{\p\over{\p\l}})
({\p\over{\p\l}}
\g^{m_5 n_5 n_2 m_6 m_7}
{\p\over{\p\l}})
({\p\over{\p\l}}
\g^{m_8 n_8 n_3 n_6 m_9}
{\p\over{\p\l}})$$
$$
({\p\over{\p\l}}
\g^{m_{10} n_{10} n_4 n_7 n_9}
{\p\over{\p\l}}) 
~{\p\over{\p B^{m_1 n_1}_1}} ... {\p\over{\p
 B^{m_{10} n_{10}}_{10}}}~]^g  $$
$$(\e{\cal T}^{-1})^{((\a\b\g))}_{[\rho_1 ...\rho_{11}]} 
{\p\over{\p\l^\a}}
{\p\over{\p\l^\b}}
{\p\over{\p\l^\g}}
{\p\over{\p C_{1\rho_1}}}
...
{\p\over{\p C_{11\rho_{11}}}} 
$$
$$\prod_{I=1}^{11}
({\p\over{\p\l^\d}} {\p\over{\p C_{I\d}}})^{K_I} 
\prod_{P=1}^{10} \prod_{R=1}^g
({\p\over{\p B_P^{pq}}}{\p\over{\p N_{R pq}}})^{L_{P,R}}
\prod_{R=1}^g ({\p\over{\p J_R}})^{M_{R}}
h(\l,N_R,J_R,C_I,B_P),$$
where the proportionality constant $c'$ can be computed as in \detc.

So as 
claimed, the final expression for the scattering amplitude is a manifestly
Lorentz-covariant function of the polarizations and momenta of the
external states.
Although this expression is complicated for arbitrary $g$-loop
amplitudes, it will be shown in the following section how this
prescription can be used to prove certain vanishing theorems.

\newsec{Amplitude Computations and Vanishing Theorems}

In this section, the amplitude prescription of section 5 will be used to
prove certain properties of closed superstring scattering amplitudes
involving massless states. In subsection (6.1), the closed
superstring vertex operator for Type IIB supergravity
states will be reviewed. In subsection (6.2),
it will be proven that massless $N$-point $g$-loop amplitudes are vanishing
whenever $N<4$ and $g>0$. In subsection (6.3), the four-point massless
one-loop amplitude will be computed. And in subsection (6.4), it will
be proven that the low-energy limit of the four-point massless amplitude
gets no perturbative contributions above one-loop.

\subsec{Type IIB supergravity vertex operator}

Just as the super-Maxwell states of the open superstring
are described by the
unintegrated vertex operator $V=\l^\a A_\a(x,\t)$ 
satisfying $QV=0$ and $\d V=Q\Omega$, the Type IIB supergravity
states of the closed superstring
are described by the unintegrated vertex operator
\eqn\closedun{V = \l^\a \bar\l^\b A_{\a\b}(x,\t,\tb)}
satisfying 
\eqn\eomc{QV = \bar Q V=0,\quad\d V = Q\Omega+\bar Q\bar\Omega}
where $\bar Q\Omega= Q\bar\Omega=0.$
The equations $QV=\bar Q V=0$ imply that
\eqn\eqc{D_\g A_{\a\b} + D_\a A_{\g\b} = \g^m_{\a\g} A_{m\b},\quad
\bar D_\g A_{\a\b} +\bar D_\b A_{\a\g} = \g^m_{\b\g} A_{\a m} }
for some superfields $A_{m\b}$ and $A_{\a m}$ where
\eqn\derivs{D_\a ={\p\over{\p\t^\a}}+\half (\g^m\t)_\a \p_m,\quad
\bar D_\a ={\p\over{\p\bar\t^\a}}+\half (\g^m\bar\t)_\a \p_m}
are the N=2 D=10 superspace derivatives.
And the gauge transformations $\d V = Q\Omega +\bar Q\bar\Omega$
where $\bar Q\Omega= Q\bar\Omega=0$ implies 
\eqn\gauc{\d A_{\a\b} = D_\a \Omega_\b + \bar D_\b \bar\Omega_\a,}
$$
\d A_{m\a} =\p_m \Omega_\a - \bar D_\a\bar\Omega_m,\quad
\d A_{\a m} =\p_m \bar\Omega_\a -  D_\a \Omega_m,$$
where 
$$D_{(\a} \bar\Omega_{\b)} = \g^m_{\a\b}\bar\Omega_m,\quad
\bar D_{(\a} \Omega_{\b)} = \g^m_{\a\b}\Omega_m. $$

In components, one can use \eqc\ and \gauc\ to gauge $A_{\a\b}(x,\t,\bar\t)$
to the form
\eqn\compc{A_{\a\b}(x,\t,\tb) = e^{ik\cdot x} [
h_{mn} (\g^m\t)_\a(\g^n\tb)_\b  + \bar\psi_m^\g (\g^m\t)_\a 
(\g^n\tb)_\b (\g_n\tb)_\g  }
$$+
\psi_n^\g (\g^m\t)_\a (\g_m\t)_\g
(\g^n\tb)_\b + F^{\g\d} (\g^m\t)_\a (\g_m\t)_\g (\g^n \tb)_\b(\g_n\tb)_\d
+ ... ] $$
where 
$$k^2 = k^m h_{mn} = k^n h_{mn} = 
k^m \bar\psi_m^\a = k^n (\g_n\bar\psi_m)_\a =0,$$
$$ k^m \psi_m^\a =
k^n (\g_n\psi_m)_\a = k_m \g^m_{\a\g} F^{\g\d} 
= k_m \g^m_{\a\d} F^{\g\d}=0,$$
and $...$ involves products of $k^m$ with $h^{mn}$, $\psi_m^\a$, 
$\bar\psi_m^\a$,
or $F^{\a\b}$.
So $A_{\a\b}(x,\t,\tb)$ describes the on-shell Type IIB supergravity
multiplet where $h_{mn}$ describes the graviton, antisymmetric tensor and
dilaton, $\psi_m^\a$ and $\bar\psi_m^\a$ describe the gravitini and
dilatini, and $F^{\a\b}$ describe the Ramond-Ramond field strengths.
Note that the $x$-independent part of 
\compc\ can be interpreted as the left-right product of two
super-Maxwell superfields $A_\a (\t) \bar A_\b(\tb)$ where 
$A_\a(x,\t) = a_m(\g^m\t) + \xi^\g (\g^m\t)_\a(\g_m\t)_\g + ...$
and 
$$h_{mn} = a_m \bar a_n,\quad \bar\psi_m^\a = a_m \bar\xi^\a,\quad 
\psi_m^\a = \xi^\a \bar a_m,\quad F^{\a\b} = \xi^\a \bar\xi^\b.$$
So one can interpret the unintegrated massless closed superstring vertex
operator of \closedun\ as the left-right product of two unintegrated
massless open superstring vertex operators using the identification
$$\l^\a\bar\l^\b A_{\a\b}(x,\t,\tb) =
e^{ik\cdot x}\l^\a A_\a(\t) \bar\l^\b \bar A_\b(\tb).$$

Just as the integrated open superstring vertex operator $U_{open}$ is
related to the unintegrated open superstring vertex operator $V_{open}$
by $Q U_{open} = \p V_{open}$, the 
integrated closed superstring vertex operator $U_{closed}$ is
related to the unintegrated closed superstring vertex operator $V_{closed}$
by $Q \bar Q U_{closed} = \p\bar\p V_{closed}$. Although one can easily
write an explicit expression for the integrated form of the
Type IIB supergravity
vertex operator \howeme\ref\laura{P.A. Grassi and L. Tamassia,
{\it Vertex Operators for Closed Superstrings}, $~~~~~$
hep-th/0405072.}, it will be more convenient to recognize that 
it is related to the left-right product of two integrated 
super-Maxwell vertex operators of \supermax.
So the integrated Type IIB supergravity vertex operator can be
expressed as 
\eqn\formvc{
U_{closed}= e^{ik\cdot x}(
\p \t^\a A_\a (\t) + 
\Pi^m A_m (\t) + d_\a W^\a (\t) +\half N^{mn} {\cal F}_{mn}(\t)) }
$$ (\bar\p \tb^\b \bar A_\b (\tb) + 
\bar\Pi^p \bar A_p (\tb) + \bar d_\b \bar W^\b (\tb) +\half \bar N^{pq} 
\bar {\cal F}_{pq}(\tb)).$$
Since the closed string graviton $h_{mn}$
is identified with the product of
$a_m \bar a_n$, the $\t=\tb=0$ component of
${\cal F}_{mn}(\t) \bar{\cal F}_{pq}(\tb)$ 
is identified with the linearized curvature tensor 
$R_{mnpq} = k_{[m} h_{n][q} k_{p]}$.

\subsec{Non-renormalization theorem}

In this subsection, the amplitude prescription of section 5 will be
used to prove that massless $N$-point $g$-loop amplitudes
vanish whenever $N<4$ and $g>0$.
For $N=0$, this implies vanishing of the cosmological constant; for
$N=1$, it implies absence of tadpoles; for $N=2$, it implies the mass
is not renormalized; and for $N=3$, it implies the coupling constant
is not renormalized. Using the arguments of \atick\martinec\ which were
summarized in the introduction, and
assuming factorization and the absence 
of unphysical divergences in the interior of moduli space,
these
non-renormalization theorems imply that superstring 
scattering amplitudes are finite order-by-order in perturbation theory.


Although
surface terms were ignored in deriving the amplitude
prescription of section 5,
it is necessary that the proof of the non-renormalization
theorem remain valid even if one includes
such surface term contributions. Otherwise, there could be divergent
surface term contributions which would invalidate the proof. For
this reason, 
one cannot assume Lorentz invariance or spacetime supersymmetry to
prove the non-renormalization theorem since the prescription of \gloop\ is
Lorentz invariant and spacetime supersymmetric only after ignoring the
surface terms. 

Fortunately, it will be possible to prove the non-renormalization theorem
using only the counting of zero modes. Since this type of argument
implies the pointwise vanishing of the integrand of the scattering
amplitude (as opposed to only implying that the integrated amplitude
vanishes), the proof remains valid if one includes the contribution
of surface terms.

On a surface of arbitrary genus, one needs 16 zero modes of $\t^\a$ and
$\tb^\a$ for the amplitude to be non-vanishing. Since the only
operators in \gloop\ containing $\t^\a$ zero modes\foot{When expressed
in terms of the free fields $(x^m,\t^\a,p_\a)$, $\Pi^m$ and $d_\a$
contain $\t$'s without derivatives which naively could contribute
$\t^\a$ zero modes. But if the supersymmetric OPE's of \oped\
are used to integrate out the non-zero worldsheet modes, the OPE's
involving $\Pi^m$ and $d_\a$ will never produce $\t^\a$ zero modes.}
are the eleven $Y_C$
picture-lowering operators and the $U_T$ vertex operators, and since
each $Y_C$ contributes a single $\t^\a$ zero mode, the $U_T$ vertex operators
must contribute at least five $\t^\a$ and five $\tb^\a$ zero modes
for the amplitude to be non-vanishing. This immediately implies that
zero-point amplitudes vanish.

For one-point amplitudes, conservation of momentum implies that
the external state must have momentum $k^m=0$. But when $k^m=0$,
the maximum number of zero modes in the vertex operator is one $\t^\a$ and
one $\tb^\a$ coming from the superfield 
$$A_{\a\b}(\t,\tb) = h_{mn} (\g^m\t)_\a (\g^n\tb)_\b.$$
All other components in the superfields appearing in the vertex operators of
\closedun\ and \formvc\
are either fermionic or involve powers of $k^m$.
So all one-point amplitudes vanish.

To prove that massless two and three-point amplitudes
vanish for non-zero $g$, one needs to count the available zero modes
of $d_\a$, as well as the zero modes of $N_{mn}$. On a genus $g$
surface, non-vanishing amplitudes require $16g$ zero modes of $d_\a$.
In addition, the number of $N_{mn}$ zero modes must be at least as large
as the number of derivatives acting on the delta functions $\d(BN)$
in the amplitude prescription. Otherwise, integration over the $N^{mn}$
zero modes will trivially vanish.

To prove the $N$-point
$g$-loop non-renormalization theorem for $N=2$ and $N=3$,
it is useful to distinguish between one-loop amplitudes and multiloop
amplitudes.
For massless $N$-point one-loop amplitudes using the prescription of \oneloop, 
there are $(N-1)$
integrated vertex operators of \formvc, each of which can either provide a
$d_\a$ zero mode or an $N_{mn}$ zero mode. So one
has at most $(N-1-M)$ $d_\a$ zero modes and $M$ $N_{mn}$ zero modes
coming from the vertex operators where $M\leq N-1$. 
Each of the nine $Z_{B_P}$ operators and 
one
$Z_J$ operator can provide a single $d_\a$ zero mode, so to get a total of
16 $d_\a$ zero modes, $b_B$ must provide at least 
\eqn\atleast{ 16- (N-1-M)-9-1 = 7-N+M }
$d_\a$ zero modes.

It is easy to verify from \stopshere\
that $b_B$ can provide a maximum of four $d_\a$
zero modes, however, the terms containing four $d_\a$ zero modes also
contain $(-1)$ $N_{mn}$ zero modes where a derivative acting on $\d(BN)$
counts as a negative $N_{mn}$ zero mode. This fact can easily be derived
from the $+4$ engineering dimension of $b_B$ where 
$[\l^\a,\t^\a,x^m,d_\a, N_{mn}]$ are defined to carry engineering
dimension $[0,\half,1,{3\over 2},2]$ and $\p^L\d(BN)$ is defined to
carry engineering dimension $-2L$. Since $(d)^4$ carries engineering dimension
$+6$, it can only appear in $b_B$ together with a term such as $\p\d(BN)$
which carries engineering dimension $-2$.

So for $N\leq 3$ and $M=0$, \atleast\ implies
that the only way to obtain 16 $d_\a$ zero
modes is if $b_B$ provides at least four $d_\a$ zero modes. But in this
case, $b_B$ contains $(-1)$ $N_{mn}$ zero modes, so the amplitudes vanish
since there are not enough $N_{mn}$ zero modes to absorb the derivatives on
$\d(BN)$. And when $M>0$, the amplitude vanishes for $N\leq 3$
since one needs more than four $d_\a$ zero modes to come from $b_B$.

For multiloop amplitudes, the argument is similar, but one now has $N$
integrated vertex operators instead of $(N-1)$. So
the vertex operators can contribute a maximum of $(N-M)$ $d_\a$
zero modes and $M$ $N_{mn}$ zero modes where $M\leq N$. And each of the $7g+3$
$Z_B$ and $g$ $Z_J$ operators can provide a single $d_\a$ zero mode.
So to get a total of $16g$ $d_\a$ zero modes, the 
$(3g-3)$ $b_B$'s must provide at least
\eqn\atlt{ 16g - (N-M) - (7g+3) -g = 8g-3-N+M} 
$d_\a$ zero modes. Since $(3g-3)$ $b_B$'s
carry engineering dimension $12g-12$, $d_\a$ carries engineering dimension
$3\over 2$, and $N_{mn}$ carries engineering dimension $+2$, the
$(3g-3)$ $b_B$'s can provide a maximum of $(8g-8)$ $d_\a$ zero modes
with no derivatives of $\d(BN)$, or $(8g-8+{4\over 3}M)$ $d_\a$
zero modes with $M$ derivatives of $\d(BN)$. 
Since 
\eqn\inequal{8g-8 +{4\over 3} M < 8g-3-N+M}
whenever $M\leq N\leq 3$, there is no way for the $(3g-3)$
$b_B$'s to provide enough $d_\a$ zero modes without providing too many
derivatives of $\d(BN)$. 

So the $N$-point multiloop non-renormalization theorem has been proven
for $N\leq 3$. Note that when $N=4$, 
\eqn\nequal{8g-8 +{4\over 3} M \geq 8g-3-N+M}
if one chooses $M=3$ or $M=4$. So 
four-point multiloop amplitudes do not need to vanish. However,
as will be shown in subsection (6.4), one can prove that
the low-energy limit of these multiloop amplitudes vanish, which 
implies that the $R^4$ term in the effective action gets no 
perturbative corrections above one loop. But before proving this,
it will be useful to see how the four-point one-loop amplitude is
reproduced in the pure spinor formalism.

\subsec{Massless four-point one-loop amplitude}

The simplest non-vanishing one-loop amplitude involves four massless
particles and can be computed using either the RNS or light-cone
GS formalism. Nevertheless, it is interesting to see how this well-known
amplitude can be derived from the super-Poincar\'e covariant 
prescription of section 5.

As discussed in \atleast, $b_B$ must provide at least $(7-N+M)$
$d_\a$ zero modes for the one-loop amplitude to be non-vanishing
where $N$ is the number of external states
and $M$ is the number of $N_{mn}$ zero modes coming from the
vertex operators. Since $b_B$ carries engineering dimension $+4$,
the only way to satisfy \atleast\ when $N=4$ is 
if $M=1$ and $b_B$ provides four 
$d_\a$ zero modes. In terms of the operators $H^{\a\b}$,
$K^{\a\b\g}$ and $L^{\a\b\g\d}$ defined in \gfour, \hfour\ and
\kfour,
the only such terms in $b_B$ are
\eqn\bbterms{{1\over 4} H^{\b\a}(Bd)_\a (Bd)_\b \p\d(BN) -
{1\over 8} K^{\g\b\a}(Bd)_\a (Bd)_\b (Bd)_\g \p^2\d(BN)}
$$ +{1\over 16}
L^{\d\g\b\a}(Bd)_\a (Bd)_\b (Bd)_\g (Bd)_\d \p^3\d(BN) $$
where $(Bd)_\a = B^{mn} (\g_{mn} d)_\a$. 

Since all $d_\a$ and $N_{mn}$ variables are used to absorb zero modes
in this correlation function,
the functional integral over the $(\t^\a,p_\a)$ variables and over the
pure spinor ghosts only contributes to the four-point one-loop amplitude
through the zero-mode integral
\eqn\zint{|\int d^{16}\t\int d^{16} d
\int [{\cal D}\l][{\cal D}N] }
$$
(-{1\over{1536}}\g_{mnp}^{\b\a}(d\g^{mnp}d) (Bd)_\a (Bd)_\b \p\d(BN) +
{1\over 8}
c_{1mn}^{\g\b\a\rho} N^{mn}d_\rho (Bd)_\a (Bd)_\b (Bd)_\g \p^2\d(BN) $$
$$-{1\over {16}}
c_{4 mnpq}^{\d\g\b\a} N^{mn} N^{pq} 
(Bd)_\a (Bd)_\b (Bd)_\g (Bd)_\d \p^3\d(BN)~ ) $$
$$\prod_{P=2}^{10} B_P^{mn} (\l \g_{mn} d) \d(B_P N) (\l d)\d(J) 
\prod_{I=1}^{11} (C_I\t)\d(C_I\l)$$
$$\l^\a A_{1\a}(\t) \prod_{T=2}^4 (d_\a W^\a_T(\t) +\half N_{mn} 
{\cal F}^{mn}_T(\t))~|^2$$
where the closed superstring vertex operators have been written as the
left-right product of open superstring vertex operators as in \formvc. 

Integrating over the constant spinors and tensors $C_{I\a}$ and
$B_P^{mn}$ and using the formula of \finalf, one finds
that the term in \zint\ which is independent of 
$c_{1mn}^{\g\b\a\rho}$ and
$c_{4 mnpq}^{\d\g\b\a}$
is proportional to
\eqn\yint{|\int d^{16}\t\int d^{16} d~
(\e {\cal T}^{-1})^{((\k_1\k_2\k_3}_{[\rho_1 ...\rho_{11}]} }
$$
(\g^{m_1 n_1 m_2 m_3 m_4})^{\k_4\k_5}
(\g^{m_5 n_5 n_2 m_6 m_7})^{\k_6\k_7}
(\g^{m_8 n_8 n_3 n_6 m_9})^{\k_8\k_9}
(\g^{m_{10} n_{10} n_4 n_7 n_9})^{\k_{10}\k_{11))}}$$
$$(d\g^{rst}d)\g_{rst}^{\s\g}(\g_{pq} d)_\s (\g_{m_1 n_1} d)_\g
(\g_{m_2 n_2} d)_{\k_2} ...
(\g_{m_{10} n_{10}} d)_{\k_{10}}
d_{\k_{11}} (\t^{\rho_1} ...\t^{\rho_{11}})$$
$$ A_{1\k_1}(\t) ~
(~d_\a W_2^\a(\t) d_\b W_3^\b(\t) {\cal F}_4^{pq}(\t)   $$
$$ +
d_\a W_3^\a(\t) d_\b W_4^\b(\t) {\cal F}_2^{pq}(\t)  +
d_\a W_4^\a(\t) d_\b W_2^\b(\t) {\cal F}_3^{pq}(\t) ~)~|^2 $$
where the proportionality constant will not be determined here.
Integrating over the $d_\a$ zero modes and 
performing gamma-matrix manipulations, one finds that this 
integral is proportional to 
\eqn\wint{|\int d^{16}\t
(\e {\cal T}^{-1})^{((\k_1\k_2\k_3))}_{[\rho_1 ...\rho_{11}]}
(\t^{\rho_1} ...\t^{\rho_{11}}) (\g_{mnpqr})_{\k_1\k_2}
A_{1\k_3}(\t) }
$$
((W_2(\t)\g^{mnp} W_3(\t)) {\cal F}_4^{qr}(\t) +
(W_3(\t)\g^{mnp} W_4(\t)) {\cal F}_2^{qr}(\t) +
(W_4(\t)\g^{mnp} W_2(\t)) {\cal F}_3^{qr}(\t) )|^2.$$

The result of \wint\ can be obtained without going through the
complicated gamma-matrix manipulations by noting that the
$\k_1\k_2\k_3$ indices on 
$(\e {\cal T}^{-1})^{((\k_1\k_2\k_3))}_{[\rho_1 ...\rho_{11}]}
(\t^{\rho_1} ...\t^{\rho_{11}})$ need to be contracted with a
$\g$-matrix traceless combination constructed from one
$A_\a$, two $W^\b$'s, and one $F_{mn}$. 
The only possible such combination is 
$(\g_{mnpqr})_{((\k_1\k_2}
A_{\k_3))} (W \g^{pqr} W) F^{mn}$. For this reason, the terms in 
\zint\ which depend on
$c_{1mn}^{\g\b\a\rho}$ and
$c_{4 mnpq}^{\d\g\b\a}$
must also give contributions proportional to \wint\ after integration
over $C_{I\a}$ and $B_P^{mn}$.

Finally, one needs to include the correlation function for the
$x^m$ variables as in \xcorr\ which gives the factor
\eqn\xfactor{\int d^{10}P |\exp
(i\pi P^2\tau + 2\pi i\sum_{T=1}^4 (k_T\cdot P) t_T) 
\prod_{T<U} E(t_T,t_U)^{k_T\cdot k_U} |^2 }
$$= (Im ~\tau)^{-5} \prod_{T<U} G(t_T,t_U)^{k_T\cdot k_U}$$
where $G(t_T,t_U) = 
|E(t_T,t_U)|^2 \exp(-2\pi (Im ~\tau)^{-1}(Im ~t_T)(Im~ t_U)).$

So up to a constant proportionality factor, the massless four-point
one-loop amplitude is 
\eqn\fouramp{{\cal A} = \int d^2 \tau (Im ~\tau)^{-5}
\int d^2 t_2\int d^2 t_3 \int d^2 t_4 \prod_{T<U} G(t_T,t_U)^{k_T\cdot k_U}}
$$|\int d^{16}\t 
(\e {\cal T}^{-1})^{((\a\b\g))}_{[\rho_1 ...\rho_{11}]}
\t^{\rho_1} ...\t^{\rho_{11}} (\g_{mnpqr})_{\b\g}
A_{1\a}(\t)$$
$$((W_2(\t)\g^{mnp} W_3(\t)) {\cal F}_4^{qr}(\t) +
(W_3(\t)\g^{mnp} W_4(\t)) {\cal F}_2^{qr}(\t) +
(W_4(\t)\g^{mnp} W_2(\t)) {\cal F}_3^{qr}(\t) )|^2.$$

One can easily check that \fouramp\ is modular invariant and has a 
structure similar to the standard expression for the four-point one-loop
amplitude. But because of the gauge superfield $A_{1\a}(\t)$ in \fouramp,
${\cal A}$ is not manifestly gauge invariant under 
\eqn\mang{\d A_{1\a} (\t) = D_\a \Omega_1(\t).}
Nevertheless, one can use properties
of pure spinors to show that the amplitude is in fact gauge invariant under
\mang.
Integrating $D_\a$ by parts in the $\int d^{16}\t$ integral, one obtains 
\eqn\oneo{\int d^{16}\t 
(\e {\cal T}^{-1})^{((\a\b\g))}_{[\rho_1 ...\rho_{11}]}
\d A_{1\a} [\t^{\rho_1} ...\t^{\rho_{11}} (\g_{mnpqr})_{\b\g}
(W_2(\t)\g^{mnp} W_3(\t)) {\cal F}_4^{qr}(\t)] }
$$= -\int d^{16}\t 
(\e {\cal T}^{-1})^{((\a\b\g))}_{[\rho_1 ...\rho_{11}]}
\Omega_1(\t) D_\a [\t^{\rho_1} ...\t^{\rho_{11}} (\g_{mnpqr})_{\b\g}
(W_2(\t)\g^{mnp} W_3(\t)) {\cal F}_4^{qr}(\t)] $$
$$ = 
\int d^{16}\t 
(\e {\cal T}^{-1})^{((\a\b\g))}_{[\rho_1 ...\rho_{11}]}
\Omega_1(\t) (\t^{\rho_1} ...\t^{\rho_{11}}) (\g_{mnpqr})_{\b\g}
D_\a [(W_2(\t)\g^{mnp} W_3(\t)) {\cal F}_4^{qr}(\t)]$$
where the identity $
(\e {\cal T}^{-1})^{((\rho_1\b\g))}_{[\rho_1 ...\rho_{11}]} =0$
was used.
To compute 
\eqn\toshow{(\g_{mnpqr})_{((\b\g}
D_{\a))} [(W_2(\t)\g^{mnp} W_3(\t)) {\cal F}_4^{qr}(\t)],}
note that Bianchi identities imply
that $D_\a W^\d ={1\over 4} (\g^{st})_\a{}^\d {\cal F}_{st}$ and
$D_\a {\cal F}^{qr} =  \p^{[q} (\g^{r]}W)_\a.$
But $(\g_r)_{\a((\b} \g^r_{\g))\d}=0$ implies that
$$(\g_{mnp}\g_{st})_{\d((\a} (\g^{mnpqr})_{\b\g))}=0\quad {\rm and}\quad
(\g_{mnpqr})_{((\b\g} (\g^r)_{\a))\d}=0,$$
so \toshow\ vanishes and ${\cal A}$ is gauge-invariant under \mang.

Since the four external vertex operators need to provide 5 $\t^\a$ and
5 $\tb^\a$ zero modes in \fouramp, this amplitude implies the presence of a
one-loop $R^4$ term in the low-energy effective action. To see this,
note that using the left-right product language of \formvc, an $R^4$ term comes
from $|F^4|^2$. In the $A_\a(\t)$, $W^\a(\t)$ and ${\cal F}_{mn}(\t)$
superfields, the $F_{mn}$
field strength
is in the component
\eqn\compfmn{A_\a(\t) = ... + (\t\g^{mnp}\t)(\g_p\t)_\a F_{mn} + ...,}
$$
W^\a(\t) = ... + (\g^{mn}\t)^\a F_{mn} + ...,\quad
{\cal F}_{mn}(\t) = F_{mn} + ...  .$$
So if $A_\a$ provides three $\t$ zero modes and each $W^\a$ provides
one $\t$ zero mode, one obtains an $|F^4|^2$ term from the 
vertex operators in \fouramp. It should be straightforward to check that the
contractions of the Lorentz indices in this $|F^4|^2$ term agrees
with the usual contractions of the one-loop $R^4$ term in the effective
action.

\subsec{Absence of multiloop $R^4$ contributions}

Although the four-point massless amplitude is expected to be non-vanishing
at all loops, there is a conjecture based on S-duality of the Type IIB
effective action that $R^4$ terms in the low-energy effective action do
not get perturbative contributions above one-loop \green. After much effort,
this conjecture was recently verified in the RNS 
formalism at two loops \iengo\phong.
As will now be shown, the multiloop prescription of section 5 
can be easily used
to prove the validity of this S-duality conjecture at all loops.

It was proven using \nequal\
that the four-point
massless multiloop amplitude vanishes unless at least three of
the four integrated vertex operators contribute an $N_{mn}$ zero mode.
Since the only operators containing $\t$ zero modes are the eleven
picture-lowering operators and the external vertex operators, the
functional integral over $\t$ zero modes in the multiloop prescription
for the four-point amplitude gives an expression of the form
\eqn\rpr{|\int d^{16}\t (\t)^{11} (d_\a W_1^\a(\t) +\half N_{pq} 
{\cal F}^{pq}_1(\t))
\prod_{T=2}^4 N_{mn} {\cal F}^{mn}_T(\t) |^2 .}

Since the external vertex operators must contribute at least 5
$\t^\a$ and $\tb^\a$ zero modes, and since $F^{mn}$ appears in the
component expansions of $W^\a$ and ${\cal F}^{mn}$ as in \compfmn,
one easily sees that there is no way to produce an $|F^4|^2$ term
which would imply an $R^4$ term in the effective action. In fact,
by examining the component expansion of the ${\cal F}_{mn}(\t)$ 
and $W^\a(\t)$
superfields, one finds that the term with fewest number of spacetime
derivatives which contributes 5 $\t$'s and
5 $\tb$'s is $|(\p F)(\p F) F^2|^2$, which would imply a
$\p^4 R^4$ contribution to the low-energy effective action.

So it has been proven that there are no multiloop contributions to
$R^4$ terms (or $\p^2 R^4$ terms) in the low-energy effective action of
the superstring. It should be noted that this proof has assumed that
the correlation function over $x^m$ does not contribute
inverse powers of 
$k^m$ which could cancel momentum factors coming from the 
$\theta$ integration in \rpr.
Although the $x^m$ correlation function does contain poles as a function
of $k^m$ when the external vertex operators collide, these poles only
contribute to non-local terms in the effective action which involve
massless propagators, and are not expected to contribute to local terms in the
effective action such as the $R^4$ term.

\vskip 15pt
{\bf Acknowledgements:} I would like to thank  
Sergey Cherkis, Michael Douglas, Michael Green,
Warren
Siegel, Mario Tonin, Brenno Carlini Vallilo,
Herman Verlinde and Edward Witten for useful discussions,
CNPq grant 300256/94-9, 
Pronex 66.2002/1998-9,
and FAPESP grant 99/12763-0
for partial financial support, and the Institute of Advanced Studies
and Funda\c{c}\~ao Instituto de F\'{\i}sica Te\'orica  
for their hospitality.

\listrefs

\end